\documentclass[
 preprint,
 amsmath,amssymb,
 aps,
prb,
]{revtex4-2}

\usepackage{graphicx}
\usepackage{dcolumn}
\usepackage{bm}
\usepackage{color}

\begin{document}

\preprint{APS/123-QED}

\title{The distinction between Ice phases VII, VIII and X}


\author{Graeme J Ackland}
 \homepage{http://www.ph.ed.ac.uk/~gja}
\affiliation{
 Centre for Science at Extreme Conditions, School of Physics and Astronomy, University of Edinburgh, EH9 3FD\\
}%
\author{Andreas Hermann}
\affiliation{
 Centre for Science at Extreme Conditions, School of Physics and Astronomy, University of Edinburgh, EH9 3FD\\
}%
\author{Kazuki Komatsu}
\affiliation{
 Geochemical Research Center, Graduate School of Science, The University of Tokyo\\
}

\author{Keishiro Yamashita, 
J.S. Loveday}
\affiliation{
 Centre for Science at Extreme Conditions,School of Physics and Astronomy, University of Edinburgh, EH9 3FD\\
}%

\date{\today}

\begin{abstract}
Ice phases VII, VIII and X are all based on a body-centered cubic arrangement of molecules, the differences coming from molecular orientation.  There is some debate as to whether these should even be considered distinct phases. The standard definition of a transition between distinct phases involves  a discontinuity in any derivative of the free energy. This can be hard to prove experimentally, and most previous theoretical works have been based on models which either have continuously differentiable free energies, or no straightforward way to determine the free energy.  Here we build a free energy model based on the common definitions of the phases ; ordered ice-VIII, orientationally disordered ice VII and proton-disordered ice X.  All transitions in this model might or might not be associated with a discontinuity in the specific heat, depending on paramaterization.  By comparing with data, we find that a  VII-X transition line exists, but it ends in a critical point hidden within the stability field of phase VIII.    If the model is correct, there is a discontinuity between VII and X, so they are separate phases.  We propose that the hidden phase boundary might be demonstrated experimentally by compression of supercooled ice VII.  
\end{abstract}

\maketitle


\section{Introduction \label{sec:level1}}

There are currently at least twenty-two known crystalline phases of ice. These phases are generally numbered with Roman numerals in the chronological order that they were discovered. The exception to this numbering system is Ice I. The stable ambient pressure form is generally called ice Ih because it has a hexagonal structure. However, like the close-packed structures, a tetrahedral network like ice I has an alternative stacking which is cubic called ice Ic and a range of stacking fault disordered structures between the end members (ice Isd).

Other ice phases exist at high pressure. In the vast majority of these phases the water forms molecules in which an oxygen atom is strongly bound to two hydrogen atoms.  With increasing pressure, a series of phases exisits with increasing packing efficiency of the crystal structure. Typically, in a given pressure range, there is a  high temperature hydrogen-disordered phase and a low temperature ordered phase with broken symmetry but similar topology. 
Very high pressure eventually breaks apart the strong molecular bonds \cite{holzapfel1972}.

A central concept for molecular ice structures is the 
Bernal-Fowler ice rules 
 \cite{bernal1933theory}. 
\begin{itemize}
    \item 
Each oxygen is covalently bonded to two hydrogen atoms with a bond angle around 105$^\circ$. 
\item
Each oxygen accepts precisely two hydrogen bonds from neighbouring hydrogen atoms.
\end{itemize} 

These two rules are insufficient to define  the orientation of the water molecules for a given arrangement of the oxygen atoms uniquely.  Therefore ices tend to have a "Pauling entropy" associated with the number of possible arrangements, consistent with the ice rules\cite{pauling1935structure}. The ice rules place long-range constraints on the number of arrangements depending on the crystal structure: a typical value of the Pauling entropy is around \cite{pauling1935structure} 3.4 J/mol/K, but exact values are not known because the various arrangements have slightly different energies\cite{herrero2014configurational}.

\begin{figure}[h]
\includegraphics[width=0.5\textwidth]{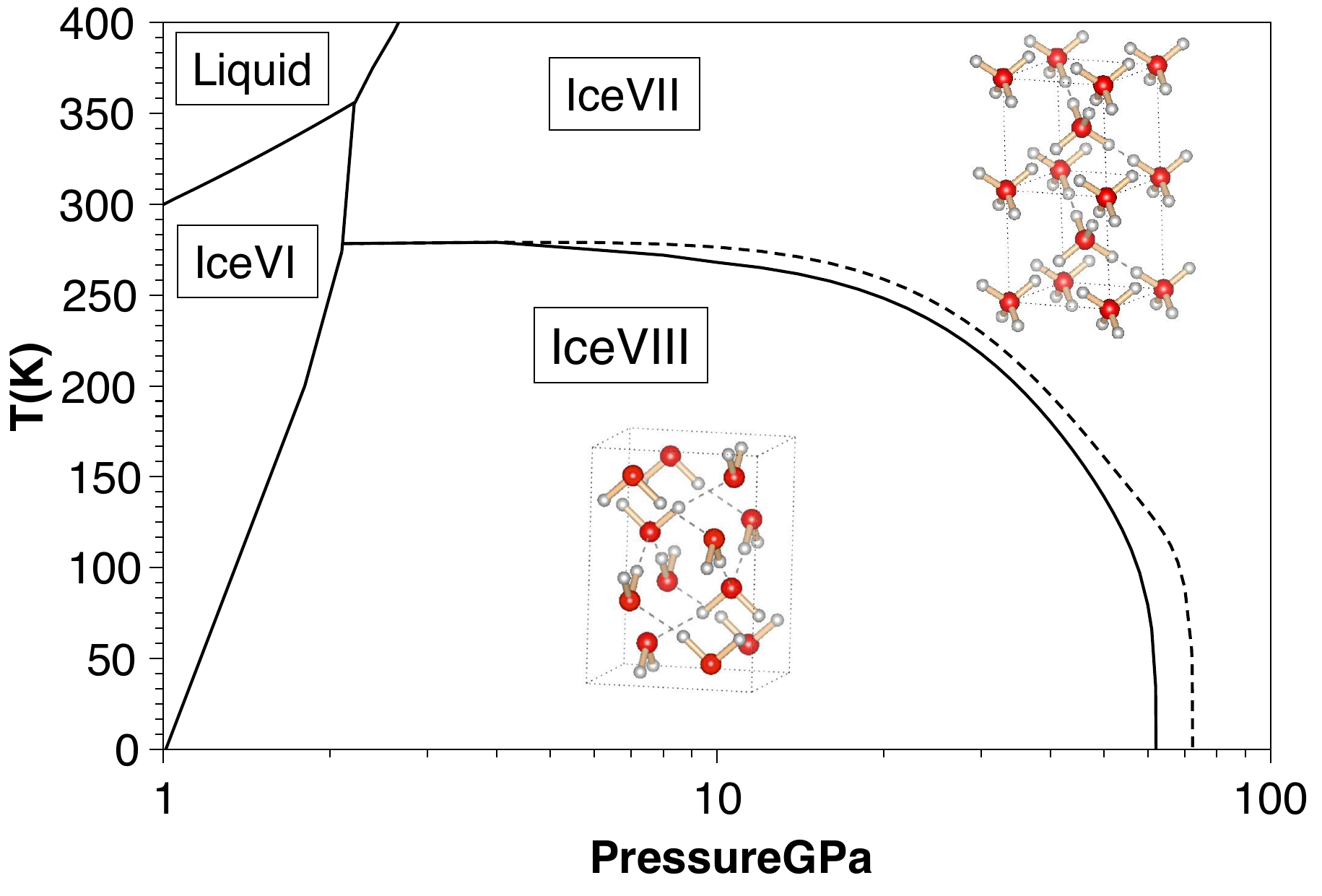}%

\caption{\label{fig:PD} The phase diagram of water. Solid/dashed lines show phase transition lines for H$_2$O/D$_2$O.}
\end{figure}

In the region of the phase diagram below 100 GPa, Ice VII occupies the largest portion \cite{pruzan2003}. Ice VII has a cubic structure in which the oxygen atoms occupy body-centred cubic (bcc) sites \cite{kuhs1984}. In the time- and space-averaged structure measured by diffraction, each oxygen atom is associated with four half-occupied hydrogen sites  $\sim$ 1 {\AA} away, arranged in a tetrahedron along the four equivalent $\langle 111 \rangle$ directions. At least at low pressures (2-$\sim$15 GPa), where all evidence points to strongly-bonded H$_2$O molecules, it is assumed ice VII obeys the ice rules.  The average structure (shown in the inset of Figure 1)  with half-occupied H sites is thus the result of the molecules' orientations fluctuating. Ice VII is thus said to be 'proton-disordered' which is written as half-atom "O-H/2...H/2-O" configurations. 
The need to maintain a tetrahedral H-bond arrangement means that each oxygen is H-bonded to only four of its eight nearest neighbours. Hence ice VII has two interpenetrating and unconnected diamond-like networks of hydrogen-bonded water \footnote{The basic structure of ice VII described here is complicated by the existence of multi-site disorder of both oxygen and hydrogen atoms \cite{kuhs1984,jorgensen1985,nelmes1998,yamashita2022,singer2005hydrogen} This phenomenon is noted for completeness but is not considered important the thrust of this paper}. 
Ice VII has a configurational entropy significantly lower than implied by the ice rules (1.94 Jmol$^{-1}$ K$^{-1}$ \cite{johari1974}\footnote{There are arithmetic errors in the values given in ref\cite{johari1974}. This value has been calculated from the given Clapeyron slopes}).

The search for non-molecular phases beyond ice VII was a significant goal  for high-pressure scientists \cite{hirsch1984,polian1984,pruzan2003,goncharov1996,goncharov1999raman,komatsu2024}. Early reports \cite{hirsch1984,polian1984} appear now to have detected other phenomena but there is now a consensus that spectroscopic studies \cite{goncharov1996,goncharov1999raman} show the transition as starting at $\sim$ 60 GPa in H$_2$O and completing at $\sim$ 120 GPa. This consensus is supported by ab-initio modelling \cite{bernasconi1998ab}.  The stability region of Ice VII extends to the first non-molecular phase, which was named ice X. 

There is a consensus that ice X is a phase in which the hydrogen atoms cease to be covalently bonded to a single oxygen atom, rather being shared between two oxygens. Consequently, the molecular character of  water is lost in Ice X.
But this proposed ice X has the same crystal symmetry as ice VII, so the structural marker of the transition from molecular to non-molecular is difficult to define conceptually and even with access to neutron diffraction data to measure experimentally \cite{komatsu2024}. As a result, a thermodynamics or crystallographic definition of the transition to ice X remains unclear. 

The phase diagram of water is shown in Figure 1. At room temperature, ice VII forms at $\sim$ 2 GPa. In the pressure range 2- $\sim$ 15 GPa (20 for D$_2$O) ice VII undergoes orientational ordering at $\sim$ 273 K to form ice VIII. In ice VIII the molecules in the two H-bonded networks order anti-parallel to form a tetragonal antiferroelectric structure.  Although ordering induces a small distortion of the ice VII structure, the principal features of the structures of ice VII and ice VIII is that they have the same basic oxygen arrangements and differ in that ice VIII has ordered H-bonds (O-H...O) and ice VII disordered H-bonds (O-H/2...H/2-O).

The ice VII/VIII boundary follows a near isotherm between 2 and $\sim$ 15  GPa. It then begins to fall with increasing P and then tends towards an isobar intersecting T=0 at 60 (80 for D$_2$O)GPa \cite{pruzan2003}. The isotherm shows no significant isotope effect, but when the transition temperature begins to fall, there is an increasing isotope effect so that for a given ordering temperature D$_2$O orders at a higher pressure than does H$_2$O.  

As has been stated, pressure has the eventual effect of breaking the covalent O-H bond leading to the formation of non-molecular ice X\cite{goncharov1996}. Whilst there is a general consensus that this happens, there is no consensus as to where the transitions from the molecular forms (Ices VII and VIII) to ice X are located. The merging of the two site (O-H/2...H/2-O) bonds of ice VII to the single site (O..H..O) bonds of ice X is likely to be a continuous process and is very difficult to detect structurally. Similarly, there is a debate as to whether ordered ice VIII transforms directly to ice X at low temperatures  or passes through a two-site ice VII state before reaching ice X as is seen in the analogous KH$_2$PO$_4$ \cite{nelmes1991}.

  Ice X has been extensively modelled with density functional theory\cite{benoit1999role,benoit2002reassigning,marques2009nature,benoit1998tunnelling,bernasconi1998ab}.  If the protons are treated classically, the appearance of a single potential well does not occur until very high pressures, with the symmetric structure showing unstable phonons until well over 110GPa\cite{benoit2002reassigning,marques2009nature}. 
At lower pressures two phonon instabilities occur.  One is  a flat band across the Brillioun zone - the local bond formation, equivalent to ice VII, and a second at the R-point leading to   the tetragonal $\textit{I}4_1/\textit{amd}$ structure of ice VIII. \cite{marques2009nature}
This splitting has a signature in the infrared, which is dominated by anharmonic effects which have been approximated by classical molecular dynamics plus Berry Phase calculations\cite{bernasconi1998ab}, which are in qualitative agreement with experiment\cite{goncharov1996} except for a pressure offset around 30 GPa, which can be attributed to the symmetrising effects of treating the protons quantum mechanically\cite{benoit1998tunnelling}.

  Despite the double-well potential persisting above 100 GPa,  simulations using path integral MD, classical MD and self-consistent phonons have revealed a centred probability distribution\cite{benoit1998tunnelling,kang2013quantum,cherubini2024quantum}.  These simulations offer three distinct scenarios: at low pressure the proton is associated with one oxygen, in an intermediate regime, it has a double-peaked probability distribution and hops/resonates between the two, at higher pressure there is a single-peaked probability distribution, when the peak separation is less than the full-width half maximum.  All this means that the width of the distribution has a maximum in the intermediate regime.
  
  A further transition to a distorted hcp structure (Pbcm) was proposed as ice XI in 1996\cite{benoit1996new}.  This was based on molecular dynamics started as bcc with only 16 molecules, which invariably forms a twinned structure due to stress generated on the Burgers path\cite{pinsook1998simulation,pinsook2000atomistic,pinsook1999calculation}, so can be interpreted as the fcc structure Ice XVIII\cite{weck2022evidence,millot2019nanosecond}.
  A superionic phase with the ice-VII like $Pn\bar3m$ symmetry was proposed \cite{cavazzoni1999superionic} 
  , but was later renamed ice XX based on experimental optical conductivity measurements consistent with the calculation \cite{prakapenka2021structure}.  Very recently, a bcc plastic-crystal form of ice VII near the melting point was reported\cite{rescigno2025observation}, in which the reorientation time of the molecules
approaches the libration period.  Unlike the proposed phase XX, the covalent H$_2$O molecules are assumed to retain their integrity in the plastic crystal.

In addition to diffraction data which probes long range order, spectroscopy is regarded as sufficient evidence for a phase transformation.  While this is pragmatic, in a disordered system such as ice VII the active modes are  localised
\cite{minceva1984,pruzan1990raman}.  Ice VII may well contain molecules with a full shell of neighbours identical to ice-VIII - indeed this configuration is energetically and hence statistically favoured and it becomes increasingly more likely as the transition to Phase VIII is approached.  

The energetics of ice VII/VIII/X can therefore be attributed to two factors: the electrostatic arrangement favouring one orientation of the water dipole over another, and the energy required to break the covalent bond.  

In thermodynamics, this leads to pressure driven competition between the cost of covalent bonding and the density gain (VIII-X transition) and temperature-driven competition between the dipole energy loss and entropy gain (VIII-VII transition).

Here we build a minimal model which uses only these two contributions to define the three distinct phases.

\section{Basic Ideas\label{sec:basis}}

In this paper we set out a thermodynamic framework in which to discuss the ice VII-VIII-X transitions.  We will work in dimensionless units: it is straightforward to rescale to fit the phase diagram and estimate parameters empirically later, but our aim is to describe the {\it all possible} behaviours of the demolecularisation transition.

We start by defining the three distinct phases from a theoretical point of view.



\begin{itemize}
\item The oxygen atoms in ice VII, VIII and X form a network equivalent to the bcc lattice.
\item The hydrogen-bonding network in ice VII, VIII and X comprises two interpenetrating lattices. 
    \item Ice VIII is a singly-degenerate antiferroelectric ordered state consistent with the ice rules. 
    \item Ice VII is a weighted average of states consistent with the ice rules.
    \item Ice X does not obey the ice rules.  Each hydrogen atom is associated with two oxygens, but there is no asymmetry between the "covalent bond" and the "hydrogen bond". 
\end{itemize}

The long-range ordering of the dipoles means that VIII has different crystal symmetry from VII and X.  All phases have vibrational entropy, but ice VII also has finite configurational entropy: VIII and X are pure states with zero configurational entropy.  Examples of these structures calculated by DFT in four-molecule cells have been used to analyse the transition mechanisms in the ice X region using graph theory\cite{li2024phase}

At finite temperature entropy will also mix the states, so our model considers the free energy of a system with a fraction $x$ of Ice X, $y$ of ice VIII and $z=(1-x-y)$ of Ice VII. 
We now consider phase transitions between the various phases.  We use the standard Ehrenfest  definition of a  phase transition, that it corresponds to a discontinuous change in a derivative of the free energy, which implies a discontinuous change in $x,y,z$

The equation of state is primarily determined by contributions which are essentially identical for all phases and not included here.
So, to describe phase transitions, we must focus on {\it excess} free energy: 
Phase VIII is the easiest phase to calculate the free energy using DFT and quasiharmonic phonons, so we will take $G_{VIII}$ as our reference free energy, defining  $G_z$ and $G_x$ as the excess free energies of phases VII and X relative to $G_{VIII}$ (and similarly for other thermodynamic quantities).  

The parameters in the model define contributions to the excess free energy.  They are defined here by:
\begin{itemize}
    \item 
    $zS_z$ is the excess entropy of Phase VII, from permutations obeying the ice rules
    \item
    $zU_z$ is the excess energy a typical Phase VII microstate compared to  Phase VIII
    \item $y^2 J_D$ is the antiferroelectric (dipole-dipole) energy of Phase VIII: the $y^2$ term arises because it depends on the amount of $y$ surrounding the site.

    \item $x(1-x) J_v$ is the enthalpy cost for mixing molecular and non-molecular microstates.
    \item
    $xU_x$ is the energy of ice X, due to breaking the covalent bonds.
    \item
    $xV_x$ is the reduced volume of Phase X compared to VII and VIII.
\end{itemize}

For simplicity, the model makes some approximations which could be relaxed.  All parameters are assumed independent of P and T, the densities of phases VII and VIII are assumed equal.  

Among these parameters, we can expect $U_x$ to have a significant isotope effect, with a smaller value in hydrogen.  This is because the higher zero-point energy makes the OH bond easier to break than the OD equivalent.

\section{Ice VII-VIII Transition\label{sec:78}}
The transition from Ice VII to VIII is first order and involves a change in symmetry. 
It is primarily driven by entropy.
A  statistical mechanical treatment of this transition considering all the permutations of hydrogen bonding which obey the ice rules and building a partition function has been done\cite{umemoto2010order}. Unfortunately, such a treatment cannot, even in principle, exhibit a phase transition because only describes a single phase.  It gives a crossover from Ice VII-like to Ice VIII-like behaviour, with no discontinuities in the thermodynamic properties.  There is no symmetry breaking because the partition function covers both phases and includes all 6 different antiferroelectric orientations of ice VIII.
Even though there is no phase transformation, the approach  allows a qualitative, and accurate, estimate of the phase boundary from the peak in specific heat.

To introduce a true phase transition, we make a mean field free energy model for VII-VIII by postulating that there is an antiferroelectric microstate with enthalpy $H_{VIII}$ and other ice-rule-obeying microstates with excess energy $U_z$.   We associate an excess entropy parameter $S_z$ with ice $VII$.   

  We can then define the free energy difference:
\begin{equation}
 G_z = G_{VIII}-G_{VII} =J_D y^2 + z U_z - z T S_z
\end{equation} 

 $J_D$,  $U_z$ and $S_z$  parameters  could be fitted to data or be calculated, the first two from DFT comparing many supercell configurations, and $S_z$ using the Pauling Entropy in ice VII or, equivalently, the latent heat.  Implicit approximations here are that $U_z$ and $J_D$ are constants, meaning that all ice VII configurations have the same density and the  electrostatic energy does not change with compression.






\subsubsection{Dynamical considerations}  

It is not possible to change the configuration of ice VII by a single molecule rotation without generating a Bjerrum defect\cite{bjerrum1952structure}, or  an OH-H$_3$O pair, either of which violate the ice rules.  Correlated defect-free reorientations within ice VII are possible if the network contains a directed loop in which each atom contributes precisely one donor and one acceptor: each molecule can simultaneously rotate switching the donor to acceptor site. However, ice-VIII has ferroelectric sublattices with no such directed loops, so it cannot be created from other ice-VII configurations, even  by such a correlated rotations.  Consequently, ice VII can be supercooled.

Ice VIII is antiferroelectric, which means that on one sublattice all molecules have their dipole moment oriented along the (001) axis.  It may form domains or twins which we assume to be large enough that they do not contribute to the extensive free energy. 

\section{Ice VIII-X Transition\label{sec:8X}}

The ice VIII-X transition line goes vertically to the T=0 axis, in accordance with the Third Law. The phase transformation is sometimes called a "quantum" phase transition and has a significant isotope effect. 

We can define a single-site quantum system representing both Phase VIII and Phase X, using a superposition of two basis states: the hydrogen covalently bonded to one or other oxygen.  We call these $\Phi_L$ and $\Phi_R$ (for left/right). These basis states represent two 
antiferroelectric domains of Phase VIII, while Phase X is a symmetric superposition   $(c_L\Phi_L+c_R\Phi_R)$ with $c_L^2 = c_R^2=1/2$ being the coefficients of the basis states.  The hydrogen position in this mixed state is centred midway between the oxygens, although this position is not necessarily a maximum in the probability density: ice X may have either a one-peaked or two-peaked symmetric wavefunction\cite{benoit1998tunnelling,benoit2002reassigning,kang2013quantum}.  The two-peaked quantum probability distribution is sometimes misleadingly referred to with words implying time dependence such as "tunnelling" or "resonant".

To describe the {\it classical} disorder ice VII in this quantum model we would have to consider multiple sites, some of which have 
$c_L^2 = 1 $ and others $c_R^2=1$.

We start by building the Hamiltonian for this system.  We include (loss of) antiferroelectric energy at the self-consistent mean field (SCF) level as in Ice VII:  $U_z$,  and an exchange interaction enthalpy $-K_x$ describing bond-breaking\footnote{Notice that the physics here  is similar to forming a chemical bond, except that here it is the proton rather than the electron shared between two atoms}.,  which increases as the overlap of the two states increases, and also incorporates the increased density of ice X compared to VII and VIII.   

\begin{eqnarray}
    \hat{H} &=&   2U_z (1- |\phi_L\rangle c_L^2\langle\phi_L| - |\phi_R\rangle c_R^2 \langle\phi_R| )   \nonumber\\
    && - |\phi_R\rangle \hat{K_x} \langle\phi_L| + |\phi_L\rangle \hat{K_x} \langle\phi_R|
    \end{eqnarray}
The first two terms represent the energy cost of breaking the ice rules, and second terms are the density gain from centring interaction.  Notice that the Hamiltonian includes $c_L$ and $c_R$ from the interaction of the state with the local field.   
The excess enthalpy of a general state   $(c_L|\Phi_L\rangle+c_R|\Phi_R\rangle)$ is then:

\begin{eqnarray}
   H({c_L})&=& 
2U_z (1-c_L^4 - c_R^4) + K_xc_L^2c_R^2 \nonumber 
\\ &=& 2c_L^2(1-c_L^2) (2U_z-K) 
\label{eq:HX}\end{eqnarray}
Where we have assumed that $c_L^2+c_R^2=1$.   We are interested in the ground state, which is the lowest energy combination as obtained by the variational principle.  From the above we obtain excess enthalpy zero for Ice VIII ($c_L=1$) and $(U_z-K)/2$ for Ice X ($c_L=1/2$). 

We approximate the pressure dependence of $K_x$ as linear: $K_x=K_0+PV_x$: physically this means that the density difference between Ice X and Ice VIII is independent of pressure and volume.   Finally, we combine the pressure independent terms to write $U_x=U_z+K_0$.  This incorporates the idea that the VIII-X transition energy includes bond-breaking as well as electrostatics.

These concepts allow us to convert the expansion coefficients $c_L$ and $c_R$ into $x$, $y$ and $z$.   Energetically, we see that the possible distinct energy minima are the mixed state (Ice X) and the two end states  $|\phi_L\rangle$ and $|\phi_R\rangle$ which represent different domains of Ice VIII.   For Ice VII, we would need to consider other, disordered, environments with localised states  $|\phi_L\rangle$ or $|\phi_R\rangle$ and no dipole interaction.

 We can define the fraction of Ice X at 100\% for $c_L=1/2$ and zero for $c_L=0$.  This allows us to write an expression for the excess free energy:
\begin{equation}
    \label{eq:Gx}
G_x  = U_x +PV_x \end{equation}

\subsubsection{Further Details}  
Since both phases are pure states, the phase line becomes vertical at T=0, obeying the Third Law.  
The $U_x$ term which we use in the model includes the antiferroelectric energy $U_z$ and bond-breaking energy $K_0$.  

This $K_0$ is the main source of isotopic effects. The OH bond has an associated vibrational frequency of more than 3300cm$^{-1}$ in H$_2$O dropping to 2900 cm$^{-1}$ in hydronium\cite{falk1957infrared}. Deuterated water will be a factor of $\sqrt{2}$ lower, 
which implies a  change in zero point energy of around 0.025eV, leading in turn to a significant isotope effect in the position of the phase line, as observed in the experiment.  Detailed quasiharmonic calculation\cite{umemoto2010order} suggest that phonon effects give a 10\% volume increase compared to the DFT enthalpy minimum, of which zero-point motion is the largest contribution.

As temperature increases, some of the disordered Ice-VII states will be mixed into the ensemble. These will give rise to diffuse scattering and broadening of spectroscopic lines.   

Even for simple choices of wavefunction and interaction, the self-consistent mean field (SCF) term is extremely complicated  and depends sensitively on the functional form chosen for the wavefunction. Rather than guess some mathematical forms for the quantum wavefunctions and the SCF, then work from an ad hoc assumption to a precise but inaccurate energy, we choose to postulate the form of the energy, Eq.\ref{eq:Gx}. In principle, the SCF and basis functions can be back-derived, but we found the solutions are neither unique nor illuminating.

\section{Ice VII-X Transition\label{sec:7X}}

We have developed well-defined thermodynamic models for the free energy differences between VII-VIII and VIII-X. This allows us to finesse the microscopic definition of the VII-X transition.  
The free energy difference between  VII and  
X is simply $\Delta  G_{VII-X} = \Delta  G_{VII-VIII} - \Delta  G_{VIII-X}.$ 

\section{Thermodynamic Model}

We now put together all the concepts discussed above to build a unified thermodynamic model for the Ice VII-VIII-X system. We use the idea of  "Pure" phases VII, VIII and X as described as above, but describe the actual system as a linear combination of the three.  A phase transition occurs when the fraction of a given component changes discontinuously: this meets the standard Ehrenfest definition of a phase transition, a discontinuity in the first derivative of the free energy (entropy or volume).
\begin{eqnarray}
 G(x,y,P,T) = G_{VIII} + xG_{x} + zG_{z}+ \nonumber \\ kT \left [x\ln x + y\ln y + z\ln z \right ]
\end{eqnarray} 

where the Roman numeral suffixed $G_{VIII}$ denotes the full, reference free energy, while $G_x$ and $G_z$ are the excess free energies above this.
The mixing entropy terms allows for some localised regions of each phase, which do not disrupt long-range order.  In principle, each excess free energy includes energy, volume and entropy terms, but following the discussion above we make some simplifying assumptions to reduce the parameter space.

\begin{itemize}
    \item Phase VII has higher entropy, $S_z$ being comprised of multiple microstates, Phases VIII and X have the same entropy
    \item Phase VIII is stabilised by the antiferroelectric dipole energy, provides by the mean field (fraction of Phase VIII) 
    \item Phase X has higher density, Phases VII and VIII are similar.
    \item It is unfavourable to mix non-molecular Phase X with VII or VIII, since causes long-range disruption to the ice rules. 
\end{itemize}

Thus we can discover phase boundaries as discontinuities in the parameterized free energy 

\begin{eqnarray}
G(P,T,x,y) &=&G_{VIII} + y^2J_D + x(1-x)J_V + \nonumber \\
&&(U_z-TS_z)(1-x-y)+ (U_x+PV_x)x \nonumber 
\\
&+kT& \left [x\ln x + y\ln y + (1-x-y)\ln(1-x-y) \right ] 
\end{eqnarray} 
\begin{eqnarray}
G(P,T) &=& min_{x,y} G(P,T,x,y) 
\end{eqnarray} 

In practice, we do this numerically, by simply calculating $G(P,T,x,y)$ on a grid of $x,y$, and pick the lowest value for $P,T$.  These calculations are implemented in a python notebook available on request.

\section{Results}

\subsection{Possible forms of the Phase diagram}

Our model allows us to calculate the thermodynamically possible behaviour for the ice VII, VIII, X phase diagram.  The key insight is that the phase boundaries between VII and VIII, and between VIII and X,  exist and meet, enclosing the ice VIII region at low T and P.   By contrast, the boundary between VII and X represents a first order phase transition which starts in the region where Ice VIII is the stable phase, and ends in a critical point.
Depending on the parameters, this critical point may be hidden inside the ice VIII region.
These  situations are illustrated in Figure \ref{fig:phase}.

\subsection{Determining the Phase Boundaries}

A first order transition implies that there may be metastability of the various phases arising from local minima in the free energy surface.  In Figure \ref{fig:Gtriangle} we show a situation close to the triple point where there are three distinct free energy minima close to the "ideal" phases VII, VIII and X.

\begin{figure}[h]
\includegraphics[width=0.33\textwidth]{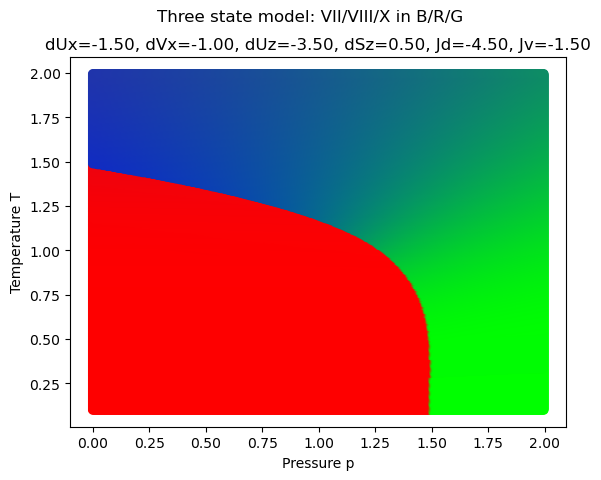}%
\includegraphics[width=0.33\textwidth]{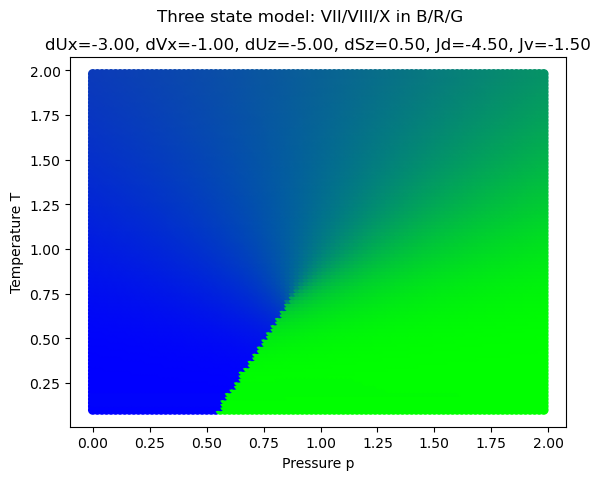}%
\includegraphics[width=0.33\textwidth]{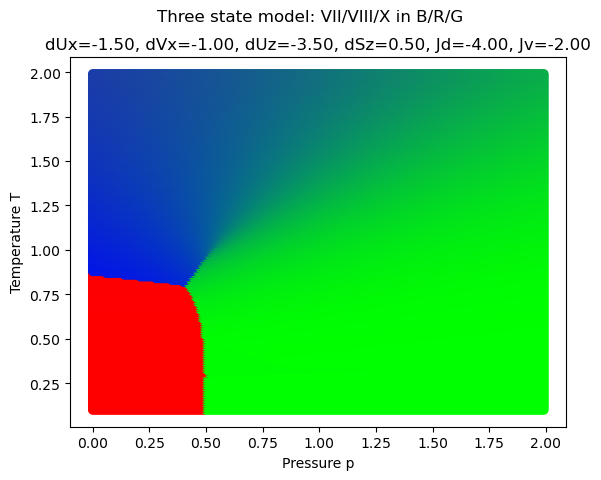}%
\caption{\label{fig:phase} RGB plot of relative amounts of VII (blue), VIII (red), X (green).  Solid lines show discontinuous changes associated with phase boundaries.
(left) VII-X Boundary and critical point hidden in Phase VIII region. (centre) 
Same model with phase VIII suppressed to reveal critical point.
(right)  VII-X Boundary and critical point outside phase VIII region, and featuring VII-VIII-X triple point.
}
\end{figure}

\begin{figure}[h]
\includegraphics[width=\textwidth]{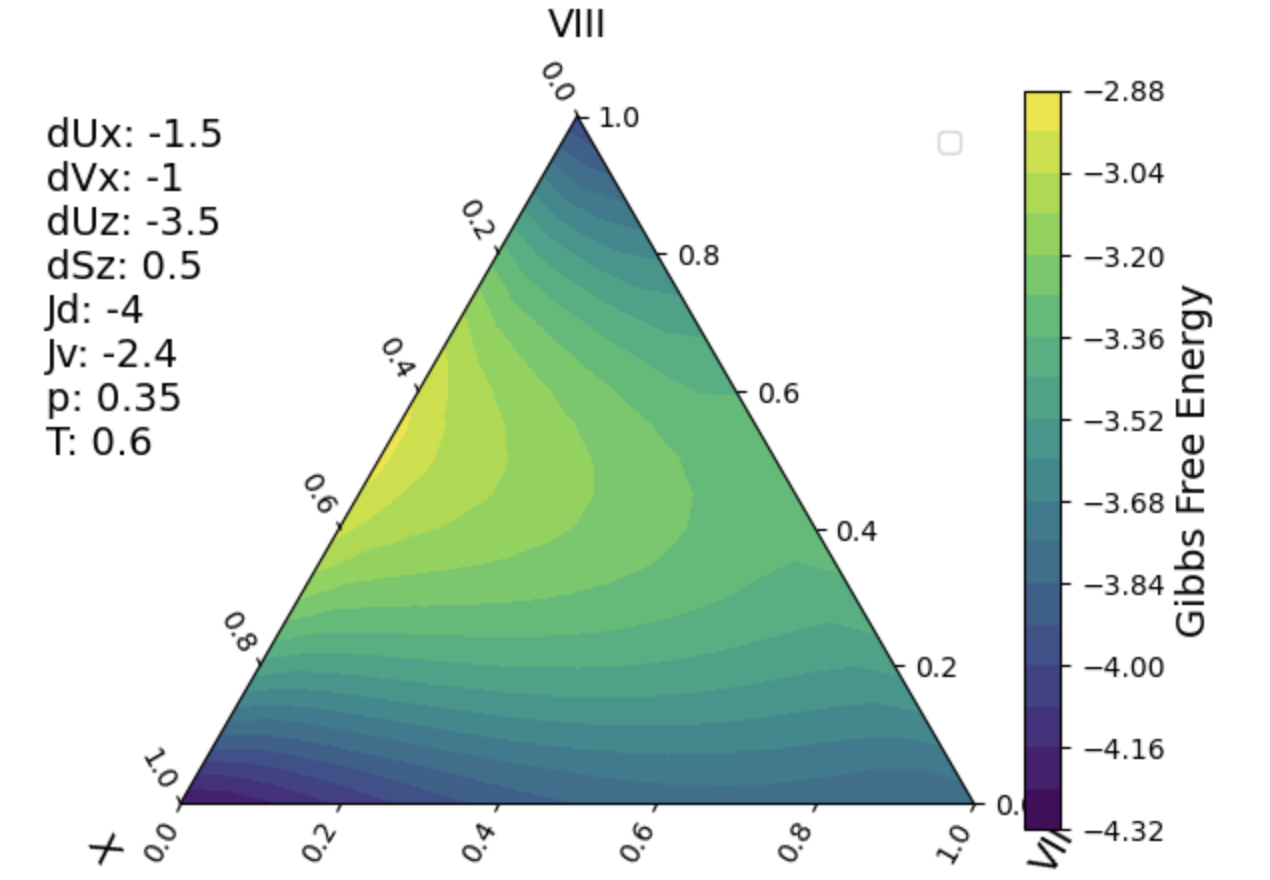} 
\caption{\label{fig:Gtriangle} G(P,T,x,y) at selected P,T and model parameters illustrating the three local minima associated with phases VII, VIII and X.  Code to investigate different settings is available online.
}
\end{figure}

Three scenarios are shown in Fig. \ref{fig:phase}. The VII-VIII boundary is close to isothermal while the VIII-X is close to isobaric. 
In the left panel parameters are chosen so that there is no VII-X phase boundary, the right panel, with different parameters, shows the VII-X boundary extending from a triple point and ending in a critical point.  The central panel  uses the same parameters as the left, except Phase VIII is suppressed by assigning it a high energy: this reveals the "hidden"  VII-X phase line and associated critical point.

\begin{figure}[h]
\includegraphics[width=0.9\textwidth]{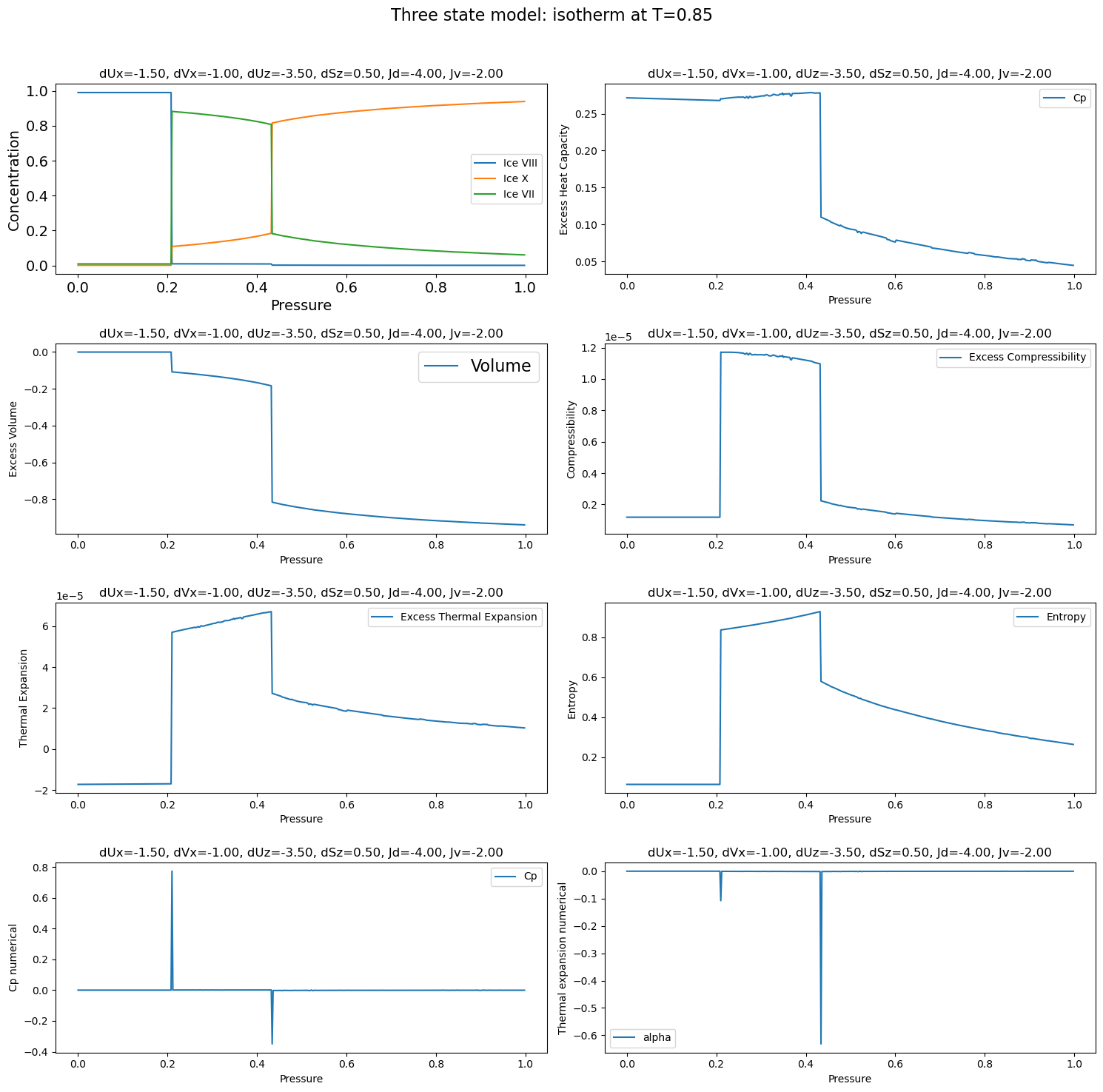}
\caption{\label{fig:Orderpar} Fractions of ice VIII, VII and X passing through two phase transitions along an isotherm at 0.85.
}
\end{figure}

A thermodynamic phase transition implies discontinuous changes in the optimal values of $(x,y,z)$
as a function of pressure and/or temperature.  In Figure \ref{fig:Orderpar}a we show the values of these quantities along a subcritical isotherm for conditions corresponding to Figure \ref{fig:phase}c.  Notice that  $(x,y,z)$ do not go rigorously to zero. 
The contribution to several associated thermodynamic quantities are also shown.

In high pressure physics, the temperature and pressure are assumed to be the independent variables.  Therefore the phase transitions are associated with discontinuous changes in the excess volume and entropy, given by:

\[ S = \frac{\partial}{\partial T_P} {\rm Min}_{x,y}(G(P,T,x,y));
V = \frac{\partial}{\partial P_T} {\rm Min}_{x,y}(G(P,T,x,y))\]

\begin{figure}[h]
\includegraphics[width=0.5\textwidth]{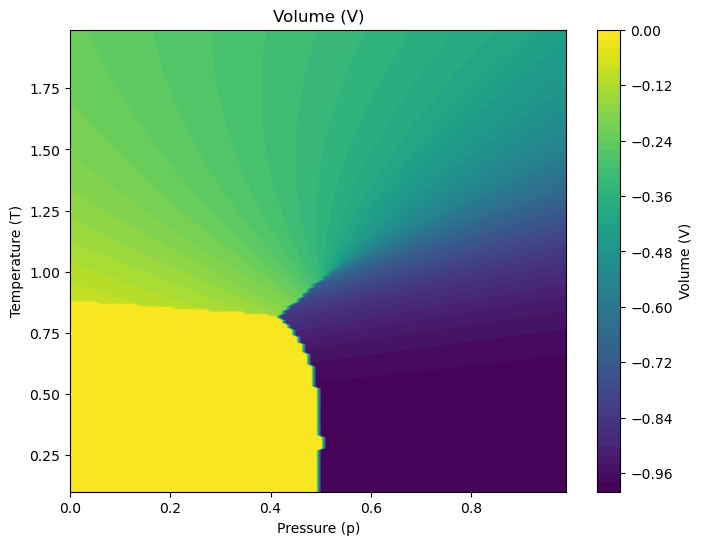}%
\includegraphics[width=0.5\textwidth]{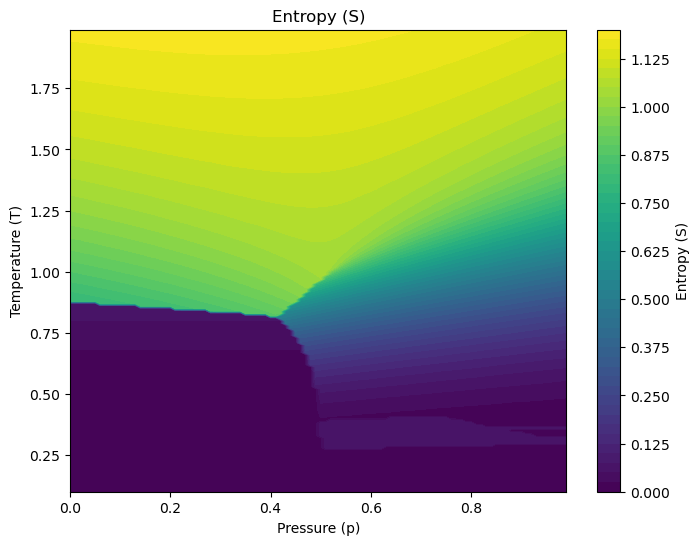}%
\caption{\label{fig:VS} (left) 2D plot of excess V vs T,P (right) 2D plot of excess S vs T,P.  The parameters here correspond to those in Figure~\ref{fig:phase}b.}
\end{figure}

These are calculated analytically and plotted in figure \ref{fig:VS}: the sharp transitions are evident.   The second derivatives, related to excess heat capacity, thermal expansion and compressibility, are plotted in figures \ref{fig:widom,fig:widomlog},
 Recall that these are excess quantities relative to Phase VIII.   
 
 Previous figures were in reduced units, so in Figure \ref{fig:asexpt} we show an {\it ab oculo} log-linear fit  to the experimental phase diagram of $dS/dP$, the thermal expansion.  The VII-VIII and VIII-X  phase boundaries appear as bright yellow lines - in fact they are discontinuities, so the dashed appearance on the VII-VIII boundary arises from how close the gridpoint is to the singularity.  Two different extremal (Widom) lines  emanate from the turning point: these are associated with the supercritical VII-X transition. 
 This illustrates that different measurements of different quantitites will give different "phase lines".  

Raman spectroscopy is a convenient probe for this region of high pressure ice.  Our model does not allow us to calculate Raman frequencies, however it is possible to make some estimates on the linewidths.  Specifically, in a disordered solid, Raman modes become localised and each molecule is in a slightly different environment and therefore vibrates with a slightly different frequency\cite{cooke2020raman,cooke2022calculating,ramsey2020localization,magduau2013identification,pena2020quantitative,howie2014phonon,frost2023isotopic,ackland2020structures}.  Strictly, these should be regarded as separate modes\cite{magduau2017infrared}, but if they are  too close to resolve experimentally they appear as a additional broadening.  There is an added complication that phase VII is itself disordered, which adds broadening $\alpha$.
Consequently, we can define an estimated linewidth from the relative amounts of the three phases:
\[ \Delta \omega \sim \alpha y + xy + yz + xz \]
Detailed calculation of linewidths is well beyond the scope of a thermodynamic model, but we observe that this increases with temperature, and as the phase boundary is approached (Fig.\ref{fig:linewidth}).

We can estimate the behaviour of the O--H stretch, which defines the transformation to phase X.   This should be related to the depth of the potential well: a double-well for phase VII and a single well for phase X.  Precisely at the transition for such a model, the force constant goes to zero as does the frequency in the harmonic approximation: within our model, the frequency depends on the enthalpy difference between Phase X and VII, which shows a minimum at the transition. In reality there is some anharmonic contribution and the system has a range of ice-X and ice-VII like environments, so it is impossible to make a more qualitative prediction.  However, we can note that the softening of the phonon occurs in both subcritical and supercritical regions, so it should not be used to determine whether there is a discontinuous thermodynamic transtion.

\section{Discussion}

\subsection{When is a phase boundary not a phase boundary?}

One cannot measure an infinite quantity in an experiment. So in practice, phase lines are determined from maxima or rapid changes in some measurable quantity.

Widom lines are maxima in thermodynamic quantities such as specific heat, thermal expansivity, compressibility and they extend beyond the VIII-X critical point. They  should not be interpreted as the phase boundary, either theoretically (because they are not discontinuities) or experimentally (because different measurable quantities have different Widom lines).

\begin{figure}[h]
\includegraphics[width=0.5\textwidth]{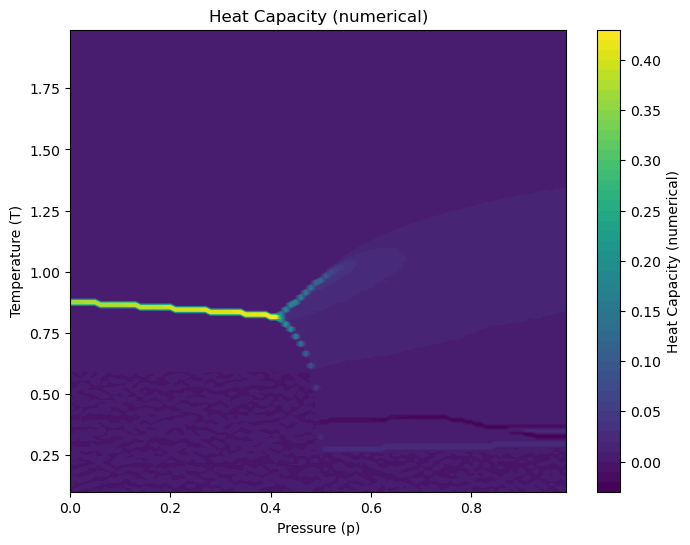}%
\includegraphics[width=0.5\textwidth]{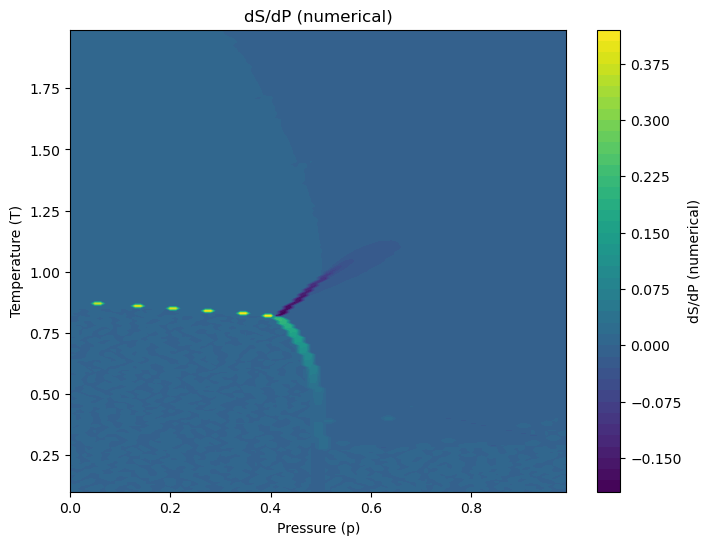}\\
\includegraphics[width=0.5\textwidth]{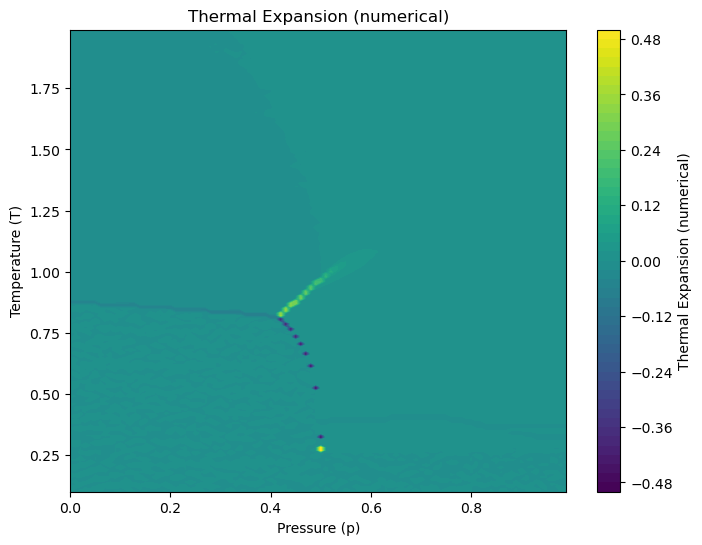}%
\includegraphics[width=0.5\textwidth]{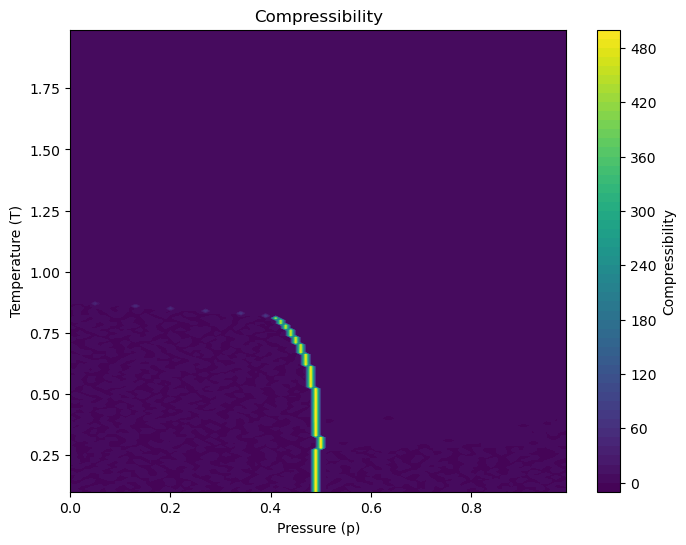}
\caption{\label{fig:widom}  Second derivatives of the Gibbs Free energy, equivalent to Heat capacity, thermal expansion and compressibility on a grid of P,T values. On a linear scale, the phase boundaries are clearly defined: they are in theory divergent, but  merely large when evaluated numerically }
\end{figure}

\begin{figure}[h]
\includegraphics[width=0.5\textwidth]{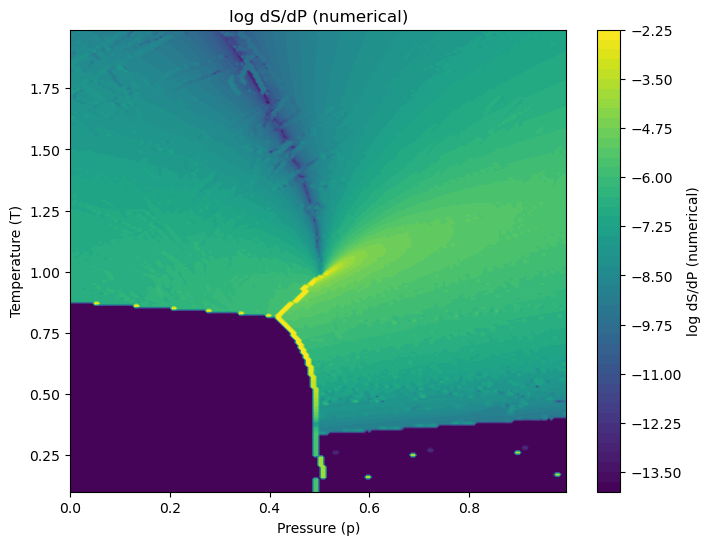}%
\includegraphics[width=0.5\textwidth]{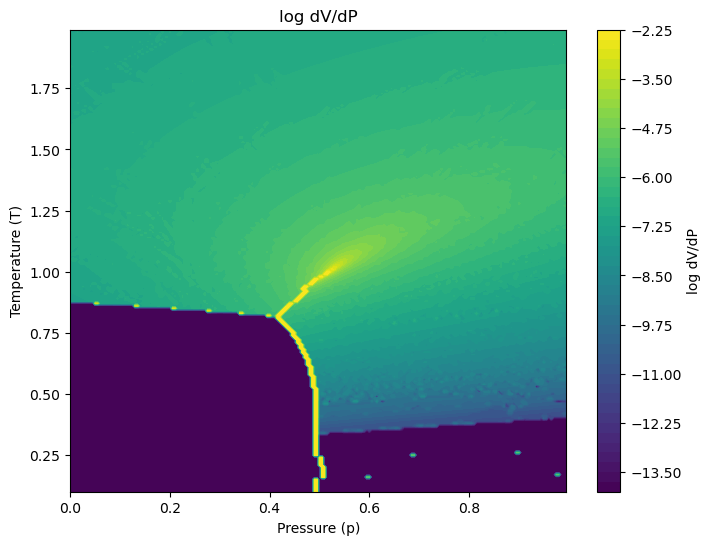}%
\caption{\label{fig:widomlog}  Second derivatives of the Gibbs Free energy, equivalent to  thermal expansion and compressibility with absolute values shown on a log scale. Extrema in these quantities, which can be seen extending beyond the critical point represent Widom Lines.  The blue line in the thermal-expansion related $dS/dP$ comes from the change from positive to negative thermal expansion. Note that the Maxwell relation  $d^2G/dPdT= \partial S/\partial P_T = \partial V/\partial T_P$ means there are only three inequivalent second derivatives. The scale is truncated so that all very low values appearing as solid dark blue at low temperature. The zero thermal expansion line is coincident with $x=z$ which we interpret as the minimum value for the Raman frequency.
}
\end{figure}

Figure \ref{fig:widom} maps out the various Widom lines extending from the phase VII/X critical point.  Notice that these lines exist even if the critical point itself lies in the stability field of Phase VIII.  A best fit of the model to the experimental data,  Fig.\ref{fig:fit} suggests that the critical point is indeed hidden. 

\subsection{Raman Modes}
One of the main probes for the putative VIII-X transition is Raman Spectroscopy.  Our thermodyanamic model cannot calculate Raman frequencies, but we can make a qualitative model based on the relative amounts of Ice VII and X.  We propose that the minimum frequency of the O-H stretch will occur when $x=z$ (at $x+z\sim 1$).   In our model, this is essentially coincident with the line of zero thermal expansion and close to an isobar. 

\subsection{Isotope Effects}

We noted above that isotopic effects are incorporated via  the parameter $U_z$.  The effect of increasing $U_z$ (deuteration) is to destabilise phase X, and move its phase boundaries to higher pressures in H$_2$O than in D$_2$O.  For some choice of parameters, it is possible to "hide" the VII-X critical point in the phase VIII stability region for D$_2$O but not for H$_2$O.  However, this requires fine tuning of the model and probably does not correspond to real ices.

\subsection{Implications for interpreting experiments on the ice VII-X transition}
In the supercritical VII-X region, it is likely that experimental probes will detect extrema in measured quantities which could be used to draw a line.  However, the position of such a line is not unique: it depends on the experimental method being used.  Most obviously from a thermodymanic viewpoint are the Widom lines seen from changes in compressibility (density), specific heat (entropy) and thermal expansion.  

Other experimental probes, such as spectroscopic measures are even more indirect.  There will be sharp changes at the phase boundary below the critical point, but even in the supercritical region there will be rapid changes in frequency with increased pressure from ice VII-like to Ice-X-like regions.  With a finite set of pressure points, a line may be drawn between the most different spectra. However, above the critical point this will not correspond to a phase transition and will not be colocated with line drawn based on other experimental data.  Changes in linewidths are also expected, but again do not signify a phase boundary.

The zero of thermal expansion corresponds to the condition $x=z$ in the supercritical region (equal amounts of ice VII and X).  It is not a Widom line, and does not resemble an extrapolation of the VII-X boundary.  However, it does emanate from the critical point, even when the critical point is hidden.  We propose that it corresponds to the minimum in the O--H stretch frequency, and explains the near-vertical phase boundary reported from that feature in Raman spectroscopy.
Like the near isobaric VIII-X boundary, this is associated with the ease of breaking the O--H bonds, and will therefore have a significant isotope effect ($U_x$). 

Consequently, the model shows that the position and shape of VII-X boundary depend on the probe used.  It is possible to define a thermodynamic boundary based on the extrapolations from the hidden critical point: such a line has a modest clapeyron slope from a balance between the higher entropy of phase VII and the higher density of phase X.  By contrast, a more local probe such as spectroscopy is insensitive to the configurational entropy, and will suggest a more vertical phase boundary.   

\subsection{Realistic parameters}

The experimental situation appears to be closest to the "hidden critical point", with no thermodynamic discontinuity to uniquely define the phase boundary.  In the absence of convincing crystallographic evidence, claims for the transition are based on spectroscopy, which the model shows will give a near-vertical "crossover" boundary, in sharp contrast with the thermodynamics boundary.

\subsection{Detectability of the hidden VII-X phase line }

 Below 10GPa the VII-VIII transition is sluggish\cite{komatsu2020anomalous}. Changing the hydrogen-bonding arrangement of ice VII cannot be attained by single molecule rotations, without breaking the ice rules.  A cooperative reorientation around a six-membered ring is required.  However, even this mechanism cannot reach the Ice VIII arrangement. 
Such supercooled Ice VII could be pressurised across the metastable phase boundary into Ice X. 
 However, the ice-X like centering of the hydrogens eliminates the ice rules.  Therefore our model implies that if metastable ice-VII is compressed across the hidden phase boundary into ice-X, the metastable ice-X will rapidly transform to ice VIII.

\section{Conclusions} 
\begin{figure}[h]
\includegraphics[width=\textwidth]{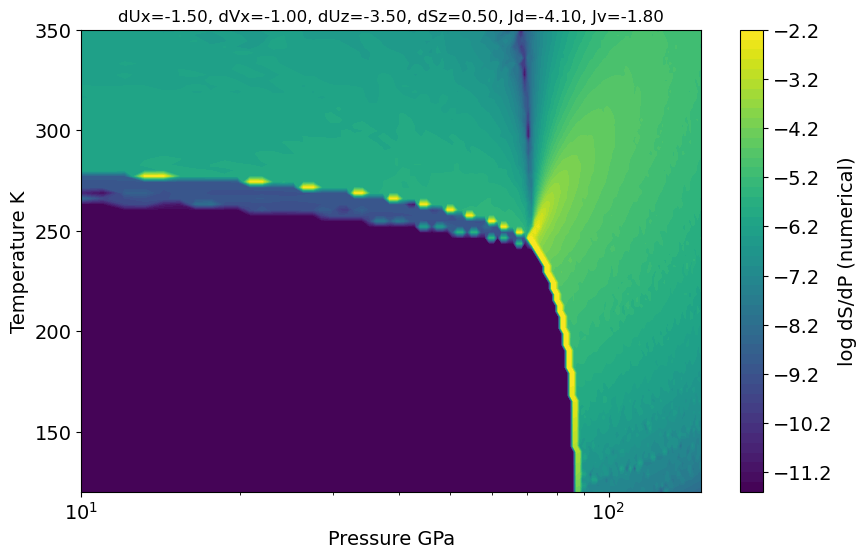}
\caption{\label{fig:asexpt}
Example of parameterisation similar to experiment.  The first order  boundary between phases VII and X ends in a critical point hidden in the ice VIII.  The plotted quantity is a surrogate for possible measurements, and the key observation is that the continuation of the VII-X boundary may follow the yellow line or the blue line.  Neither is a discontinity, but either could easily be interpreted as {\bf the} phase boundary.}\label{fig:fit}
\end{figure}

\begin{figure}[h]
\includegraphics[width=0.5\textwidth]{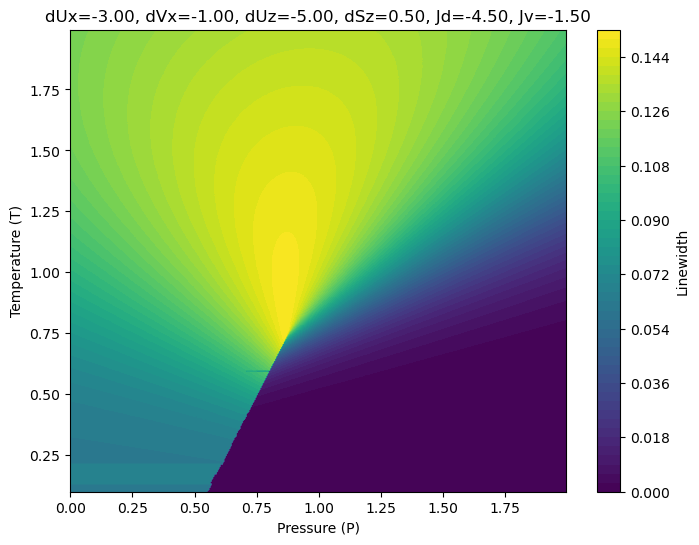}%
\caption{Plot of the function $\alpha y + xy + xz + yz$ which is related to the estimated Raman linewidth.  The important feature of this is the increased linewidth in the supercritical region. 
 If this is interpreted as the ice VII-X transition, then the line is close to an isobar, in contrast to the thermodynamic boundary and the Widom lines which have a modest, positive Clapeyron slope. 
\label{fig:linewidth}}
\end{figure}

This paper set out to determine whether "Phase X" should be regarded as a phase distinct from Ice VII.  We made a model with precise definitions of these two phases (and Ice VIII) based on theoretical models.  Within the model, Ice VII and Ice X are distinct thermodynamic phases separated by a phase boundary ending in a critical point.  Comparison with experimental results suggests that the critical point lies in the region of Phase VIII and by implication the reported measurements lie in the supercritical region(Fig. \ref{fig:fit}.

Historically, theoretical predictions have not been deemed sufficient to allocate a Roman Numeral to an ice phase. Whether the existing observations of Ice VII/X in the supercritical region constitute observations of two distinct phases is essentially a matter of definition rather than fact.

\bibliography{apssamp,icerefs}

\begin{thebibliography}{52}%
\makeatletter
\providecommand \@ifxundefined [1]{%
 \@ifx{#1\undefined}
}%
\providecommand \@ifnum [1]{%
 \ifnum #1\expandafter \@firstoftwo
 \else \expandafter \@secondoftwo
 \fi
}%
\providecommand \@ifx [1]{%
 \ifx #1\expandafter \@firstoftwo
 \else \expandafter \@secondoftwo
 \fi
}%
\providecommand \natexlab [1]{#1}%
\providecommand \enquote  [1]{``#1''}%
\providecommand \bibnamefont  [1]{#1}%
\providecommand \bibfnamefont [1]{#1}%
\providecommand \citenamefont [1]{#1}%
\providecommand \href@noop [0]{\@secondoftwo}%
\providecommand \href [0]{\begingroup \@sanitize@url \@href}%
\providecommand \@href[1]{\@@startlink{#1}\@@href}%
\providecommand \@@href[1]{\endgroup#1\@@endlink}%
\providecommand \@sanitize@url [0]{\catcode `\\12\catcode `\$12\catcode `\&12\catcode `\#12\catcode `\^12\catcode `\_12\catcode `\%12\relax}%
\providecommand \@@startlink[1]{}%
\providecommand \@@endlink[0]{}%
\providecommand \url  [0]{\begingroup\@sanitize@url \@url }%
\providecommand \@url [1]{\endgroup\@href {#1}{\urlprefix }}%
\providecommand \urlprefix  [0]{URL }%
\providecommand \Eprint [0]{\href }%
\providecommand \doibase [0]{https://doi.org/}%
\providecommand \selectlanguage [0]{\@gobble}%
\providecommand \bibinfo  [0]{\@secondoftwo}%
\providecommand \bibfield  [0]{\@secondoftwo}%
\providecommand \translation [1]{[#1]}%
\providecommand \BibitemOpen [0]{}%
\providecommand \bibitemStop [0]{}%
\providecommand \bibitemNoStop [0]{.\EOS\space}%
\providecommand \EOS [0]{\spacefactor3000\relax}%
\providecommand \BibitemShut  [1]{\csname bibitem#1\endcsname}%
\let\auto@bib@innerbib\@empty
\bibitem [{\citenamefont {Holzapfel}(1972)}]{holzapfel1972}%
  \BibitemOpen
  \bibfield  {author} {\bibinfo {author} {\bibfnamefont {W.~B.}\ \bibnamefont {Holzapfel}},\ }\bibfield  {title} {\bibinfo {title} {On the {Symmetry} of the {Hydrogen} {Bonds} in {Ice} {VII}},\ }\href {https://doi.org/10.1063/1.1677221} {\bibfield  {journal} {\bibinfo  {journal} {The Journal of Chemical Physics}\ }\textbf {\bibinfo {volume} {56}},\ \bibinfo {pages} {712} (\bibinfo {year} {1972})}\BibitemShut {NoStop}%
\bibitem [{\citenamefont {Bernal}\ \emph {et~al.}(1933)\citenamefont {Bernal}, \citenamefont {Fowler} \emph {et~al.}}]{bernal1933theory}%
  \BibitemOpen
  \bibfield  {author} {\bibinfo {author} {\bibfnamefont {J.~D.}\ \bibnamefont {Bernal}}, \bibinfo {author} {\bibfnamefont {R.~H.}\ \bibnamefont {Fowler}}, \emph {et~al.},\ }\bibfield  {title} {\bibinfo {title} {A theory of water and ionic solution, with particular reference to hydrogen and hydroxyl ions},\ }\href@noop {} {\bibfield  {journal} {\bibinfo  {journal} {J. Chem. Phys}\ }\textbf {\bibinfo {volume} {1}},\ \bibinfo {pages} {515} (\bibinfo {year} {1933})}\BibitemShut {NoStop}%
\bibitem [{\citenamefont {Pauling}(1935)}]{pauling1935structure}%
  \BibitemOpen
  \bibfield  {author} {\bibinfo {author} {\bibfnamefont {L.}~\bibnamefont {Pauling}},\ }\bibfield  {title} {\bibinfo {title} {The structure and entropy of ice and of other crystals with some randomness of atomic arrangement},\ }\href@noop {} {\bibfield  {journal} {\bibinfo  {journal} {Journal of the American Chemical Society}\ }\textbf {\bibinfo {volume} {57}},\ \bibinfo {pages} {2680} (\bibinfo {year} {1935})}\BibitemShut {NoStop}%
\bibitem [{\citenamefont {Herrero}\ and\ \citenamefont {Ram{\'\i}rez}(2014)}]{herrero2014configurational}%
  \BibitemOpen
  \bibfield  {author} {\bibinfo {author} {\bibfnamefont {C.~P.}\ \bibnamefont {Herrero}}\ and\ \bibinfo {author} {\bibfnamefont {R.}~\bibnamefont {Ram{\'\i}rez}},\ }\bibfield  {title} {\bibinfo {title} {Configurational entropy of hydrogen-disordered ice polymorphs},\ }\href@noop {} {\bibfield  {journal} {\bibinfo  {journal} {The Journal of Chemical Physics}\ }\textbf {\bibinfo {volume} {140}} (\bibinfo {year} {2014})}\BibitemShut {NoStop}%
\bibitem [{\citenamefont {Pruzan}\ \emph {et~al.}(2003)\citenamefont {Pruzan}, \citenamefont {Chervin}, \citenamefont {Wolanin}, \citenamefont {Canny}, \citenamefont {Gauthier},\ and\ \citenamefont {Hanfland}}]{pruzan2003}%
  \BibitemOpen
  \bibfield  {author} {\bibinfo {author} {\bibfnamefont {P.}~\bibnamefont {Pruzan}}, \bibinfo {author} {\bibfnamefont {J.~C.}\ \bibnamefont {Chervin}}, \bibinfo {author} {\bibfnamefont {E.}~\bibnamefont {Wolanin}}, \bibinfo {author} {\bibfnamefont {B.}~\bibnamefont {Canny}}, \bibinfo {author} {\bibfnamefont {M.}~\bibnamefont {Gauthier}},\ and\ \bibinfo {author} {\bibfnamefont {M.}~\bibnamefont {Hanfland}},\ }\bibfield  {title} {\bibinfo {title} {Phase diagram of ice in the {VII}–{VIII}–{X} domain. {Vibrational} and structural data for strongly compressed ice {VIII}},\ }\href {https://doi.org/10.1002/jrs.1039} {\bibfield  {journal} {\bibinfo  {journal} {Journal of Raman Spectroscopy}\ }\textbf {\bibinfo {volume} {34}},\ \bibinfo {pages} {591} (\bibinfo {year} {2003})}\BibitemShut {NoStop}%
\bibitem [{\citenamefont {Kuhs}\ \emph {et~al.}(1984)\citenamefont {Kuhs}, \citenamefont {Finney}, \citenamefont {Vettier},\ and\ \citenamefont {Bliss}}]{kuhs1984}%
  \BibitemOpen
  \bibfield  {author} {\bibinfo {author} {\bibfnamefont {W.~F.}\ \bibnamefont {Kuhs}}, \bibinfo {author} {\bibfnamefont {J.~L.}\ \bibnamefont {Finney}}, \bibinfo {author} {\bibfnamefont {C.}~\bibnamefont {Vettier}},\ and\ \bibinfo {author} {\bibfnamefont {D.~V.}\ \bibnamefont {Bliss}},\ }\bibfield  {title} {\bibinfo {title} {Structure and hydrogen ordering in ices {VI}, {VII}, and {VIII} by neutron powder diffraction},\ }\href {https://doi.org/10.1063/1.448109} {\bibfield  {journal} {\bibinfo  {journal} {The Journal of Chemical Physics}\ }\textbf {\bibinfo {volume} {81}},\ \bibinfo {pages} {3612} (\bibinfo {year} {1984})}\BibitemShut {NoStop}%
\bibitem [{Note1()}]{Note1}%
  \BibitemOpen
  \bibinfo {note} {The basic structure of ice VII described here is complicated by the existence of multi-site disorder of both oxygen and hydrogen atoms \cite {kuhs1984,jorgensen1985,nelmes1998,yamashita2022,singer2005hydrogen} This phenomenon is noted for completeness but is not considered important the thrust of this paper}\BibitemShut {NoStop}%
\bibitem [{\citenamefont {Johari}\ \emph {et~al.}(1974)\citenamefont {Johari}, \citenamefont {Lavergne},\ and\ \citenamefont {Whalley}}]{johari1974}%
  \BibitemOpen
  \bibfield  {author} {\bibinfo {author} {\bibfnamefont {G.~P.}\ \bibnamefont {Johari}}, \bibinfo {author} {\bibfnamefont {A.}~\bibnamefont {Lavergne}},\ and\ \bibinfo {author} {\bibfnamefont {E.}~\bibnamefont {Whalley}},\ }\bibfield  {title} {\bibinfo {title} {Dielectric properties of ice {VII} and {VIII} and the phase boundary between ice {VI} and {VII}},\ }\href {https://doi.org/10.1063/1.1681733} {\bibfield  {journal} {\bibinfo  {journal} {The Journal of Chemical Physics}\ }\textbf {\bibinfo {volume} {61}},\ \bibinfo {pages} {4292} (\bibinfo {year} {1974})}\BibitemShut {NoStop}%
\bibitem [{Note2()}]{Note2}%
  \BibitemOpen
  \bibinfo {note} {There are arithmetic errors in the values given in ref\cite {johari1974}. This value has been calculated from the given Clapeyron slopes}\BibitemShut {NoStop}%
\bibitem [{\citenamefont {{K.R.Hirsch}}\ and\ \citenamefont {{W.B.Holzapfel}}(1984)}]{hirsch1984}%
  \BibitemOpen
  \bibfield  {author} {\bibinfo {author} {\bibnamefont {{K.R.Hirsch}}}\ and\ \bibinfo {author} {\bibnamefont {{W.B.Holzapfel}}},\ }\bibfield  {title} {\bibinfo {title} {Symmetric {Hydrogen} {Bonds} in {Ice} {X}},\ }\href@noop {} {\bibfield  {journal} {\bibinfo  {journal} {Physics Letters A}\ }\textbf {\bibinfo {volume} {101}},\ \bibinfo {pages} {142} (\bibinfo {year} {1984})}\BibitemShut {NoStop}%
\bibitem [{\citenamefont {Polian}\ and\ \citenamefont {Grimsditch}(1984)}]{polian1984}%
  \BibitemOpen
  \bibfield  {author} {\bibinfo {author} {\bibfnamefont {A.}~\bibnamefont {Polian}}\ and\ \bibinfo {author} {\bibfnamefont {M.}~\bibnamefont {Grimsditch}},\ }\bibfield  {title} {\bibinfo {title} {New {High}-{Pressure} {Phase} of \$\{{\textbackslash}mathrm\{{H}\}\}\_\{2\}\${O}: {Ice} {X}},\ }\href {https://doi.org/10.1103/PhysRevLett.52.1312} {\bibfield  {journal} {\bibinfo  {journal} {Physical Review Letters}\ }\textbf {\bibinfo {volume} {52}},\ \bibinfo {pages} {1312} (\bibinfo {year} {1984})},\ \bibinfo {note} {publisher: American Physical Society}\BibitemShut {NoStop}%
\bibitem [{\citenamefont {Goncharov}\ \emph {et~al.}(1996)\citenamefont {Goncharov}, \citenamefont {Struzhkin}, \citenamefont {Somayazulu}, \citenamefont {Hemley},\ and\ \citenamefont {Mao}}]{goncharov1996}%
  \BibitemOpen
  \bibfield  {author} {\bibinfo {author} {\bibfnamefont {A.~F.}\ \bibnamefont {Goncharov}}, \bibinfo {author} {\bibfnamefont {V.~V.}\ \bibnamefont {Struzhkin}}, \bibinfo {author} {\bibfnamefont {M.~S.}\ \bibnamefont {Somayazulu}}, \bibinfo {author} {\bibfnamefont {R.~J.}\ \bibnamefont {Hemley}},\ and\ \bibinfo {author} {\bibfnamefont {H.~K.}\ \bibnamefont {Mao}},\ }\bibfield  {title} {\bibinfo {title} {Compression of {Ice} to 210 {Gigapascals}: {Infrared} {Evidence} for a {Symmetric} {Hydrogen}-{Bonded} {Phase}},\ }\href {https://www.jstor.org/stable/2890413} {\bibfield  {journal} {\bibinfo  {journal} {Science}\ }\textbf {\bibinfo {volume} {273}},\ \bibinfo {pages} {218} (\bibinfo {year} {1996})},\ \bibinfo {note} {publisher: American Association for the Advancement of Science}\BibitemShut {NoStop}%
\bibitem [{\citenamefont {Goncharov}\ \emph {et~al.}(1999)\citenamefont {Goncharov}, \citenamefont {Struzhkin}, \citenamefont {Mao},\ and\ \citenamefont {Hemley}}]{goncharov1999raman}%
  \BibitemOpen
  \bibfield  {author} {\bibinfo {author} {\bibfnamefont {A.~F.}\ \bibnamefont {Goncharov}}, \bibinfo {author} {\bibfnamefont {V.~V.}\ \bibnamefont {Struzhkin}}, \bibinfo {author} {\bibfnamefont {H.-k.}\ \bibnamefont {Mao}},\ and\ \bibinfo {author} {\bibfnamefont {R.~J.}\ \bibnamefont {Hemley}},\ }\bibfield  {title} {\bibinfo {title} {Raman spectroscopy of dense h2o and the transition to symmetric hydrogen bonds},\ }\href@noop {} {\bibfield  {journal} {\bibinfo  {journal} {Physical review letters}\ }\textbf {\bibinfo {volume} {83}},\ \bibinfo {pages} {1998} (\bibinfo {year} {1999})}\BibitemShut {NoStop}%
\bibitem [{\citenamefont {Komatsu}\ \emph {et~al.}(2024)\citenamefont {Komatsu}, \citenamefont {Hattori}, \citenamefont {Klotz}, \citenamefont {Machida}, \citenamefont {Yamashita}, \citenamefont {Ito}, \citenamefont {Kobayashi}, \citenamefont {Irifune}, \citenamefont {Shinmei}, \citenamefont {Sano-Furukawa},\ and\ \citenamefont {Kagi}}]{komatsu2024}%
  \BibitemOpen
  \bibfield  {author} {\bibinfo {author} {\bibfnamefont {K.}~\bibnamefont {Komatsu}}, \bibinfo {author} {\bibfnamefont {T.}~\bibnamefont {Hattori}}, \bibinfo {author} {\bibfnamefont {S.}~\bibnamefont {Klotz}}, \bibinfo {author} {\bibfnamefont {S.}~\bibnamefont {Machida}}, \bibinfo {author} {\bibfnamefont {K.}~\bibnamefont {Yamashita}}, \bibinfo {author} {\bibfnamefont {H.}~\bibnamefont {Ito}}, \bibinfo {author} {\bibfnamefont {H.}~\bibnamefont {Kobayashi}}, \bibinfo {author} {\bibfnamefont {T.}~\bibnamefont {Irifune}}, \bibinfo {author} {\bibfnamefont {T.}~\bibnamefont {Shinmei}}, \bibinfo {author} {\bibfnamefont {A.}~\bibnamefont {Sano-Furukawa}},\ and\ \bibinfo {author} {\bibfnamefont {H.}~\bibnamefont {Kagi}},\ }\bibfield  {title} {\bibinfo {title} {Hydrogen bond symmetrisation in {D2O} ice observed by neutron diffraction},\ }\href {https://doi.org/10.1038/s41467-024-48932-8} {\bibfield  {journal} {\bibinfo  {journal} {Nature Communications}\ }\textbf {\bibinfo {volume} {15}},\ \bibinfo {pages} {5100}
  (\bibinfo {year} {2024})}\BibitemShut {NoStop}%
\bibitem [{\citenamefont {Bernasconi}\ \emph {et~al.}(1998)\citenamefont {Bernasconi}, \citenamefont {Silvestrelli},\ and\ \citenamefont {Parrinello}}]{bernasconi1998ab}%
  \BibitemOpen
  \bibfield  {author} {\bibinfo {author} {\bibfnamefont {M.}~\bibnamefont {Bernasconi}}, \bibinfo {author} {\bibfnamefont {P.}~\bibnamefont {Silvestrelli}},\ and\ \bibinfo {author} {\bibfnamefont {M.}~\bibnamefont {Parrinello}},\ }\bibfield  {title} {\bibinfo {title} {Ab initio infrared absorption study of the hydrogen-bond symmetrization in ice},\ }\href@noop {} {\bibfield  {journal} {\bibinfo  {journal} {Physical review letters}\ }\textbf {\bibinfo {volume} {81}},\ \bibinfo {pages} {1235} (\bibinfo {year} {1998})}\BibitemShut {NoStop}%
\bibitem [{\citenamefont {Nelmes}\ \emph {et~al.}(1991)\citenamefont {Nelmes}, \citenamefont {Mcmahon}, \citenamefont {Piltz},\ and\ \citenamefont {Wright}}]{nelmes1991}%
  \BibitemOpen
  \bibfield  {author} {\bibinfo {author} {\bibfnamefont {R.}~\bibnamefont {Nelmes}}, \bibinfo {author} {\bibfnamefont {M.}~\bibnamefont {Mcmahon}}, \bibinfo {author} {\bibfnamefont {R.}~\bibnamefont {Piltz}},\ and\ \bibinfo {author} {\bibfnamefont {N.}~\bibnamefont {Wright}},\ }\bibfield  {title} {\bibinfo {title} {High-{Pressure} {Neutron}-{Diffraction} {Studies} of {Kh2po4}-{Type} {Phase}-{Transitions} as {Tc} {Tends} to 0k},\ }\href {https://doi.org/10.1080/00150199108209465} {\bibfield  {journal} {\bibinfo  {journal} {Ferroelectrics}\ }\textbf {\bibinfo {volume} {124}},\ \bibinfo {pages} {355} (\bibinfo {year} {1991})},\ \bibinfo {note} {place: Reading Publisher: Gordon Breach Sci Publ Ltd WOS:A1991HF28800062}\BibitemShut {NoStop}%
\bibitem [{\citenamefont {Benoit}\ \emph {et~al.}(1999)\citenamefont {Benoit}, \citenamefont {Marx},\ and\ \citenamefont {Parrinello}}]{benoit1999role}%
  \BibitemOpen
  \bibfield  {author} {\bibinfo {author} {\bibfnamefont {M.}~\bibnamefont {Benoit}}, \bibinfo {author} {\bibfnamefont {D.}~\bibnamefont {Marx}},\ and\ \bibinfo {author} {\bibfnamefont {M.}~\bibnamefont {Parrinello}},\ }\bibfield  {title} {\bibinfo {title} {The role of quantum effects and ionic defects in high-density ice},\ }\href@noop {} {\bibfield  {journal} {\bibinfo  {journal} {Solid State Ionics}\ }\textbf {\bibinfo {volume} {125}},\ \bibinfo {pages} {23} (\bibinfo {year} {1999})}\BibitemShut {NoStop}%
\bibitem [{\citenamefont {Benoit}\ \emph {et~al.}(2002)\citenamefont {Benoit}, \citenamefont {Romero},\ and\ \citenamefont {Marx}}]{benoit2002reassigning}%
  \BibitemOpen
  \bibfield  {author} {\bibinfo {author} {\bibfnamefont {M.}~\bibnamefont {Benoit}}, \bibinfo {author} {\bibfnamefont {A.~H.}\ \bibnamefont {Romero}},\ and\ \bibinfo {author} {\bibfnamefont {D.}~\bibnamefont {Marx}},\ }\bibfield  {title} {\bibinfo {title} {Reassigning hydrogen-bond centering in dense ice},\ }\href@noop {} {\bibfield  {journal} {\bibinfo  {journal} {Physical review letters}\ }\textbf {\bibinfo {volume} {89}},\ \bibinfo {pages} {145501} (\bibinfo {year} {2002})}\BibitemShut {NoStop}%
\bibitem [{\citenamefont {Marqu{\'e}s}\ \emph {et~al.}(2009)\citenamefont {Marqu{\'e}s}, \citenamefont {Ackland},\ and\ \citenamefont {Loveday}}]{marques2009nature}%
  \BibitemOpen
  \bibfield  {author} {\bibinfo {author} {\bibfnamefont {M.}~\bibnamefont {Marqu{\'e}s}}, \bibinfo {author} {\bibfnamefont {G.~J.}\ \bibnamefont {Ackland}},\ and\ \bibinfo {author} {\bibfnamefont {J.~S.}\ \bibnamefont {Loveday}},\ }\bibfield  {title} {\bibinfo {title} {Nature and stability of ice x},\ }\href@noop {} {\bibfield  {journal} {\bibinfo  {journal} {High Pressure Research}\ }\textbf {\bibinfo {volume} {29}},\ \bibinfo {pages} {208} (\bibinfo {year} {2009})}\BibitemShut {NoStop}%
\bibitem [{\citenamefont {Benoit}\ \emph {et~al.}(1998)\citenamefont {Benoit}, \citenamefont {Marx},\ and\ \citenamefont {Parrinello}}]{benoit1998tunnelling}%
  \BibitemOpen
  \bibfield  {author} {\bibinfo {author} {\bibfnamefont {M.}~\bibnamefont {Benoit}}, \bibinfo {author} {\bibfnamefont {D.}~\bibnamefont {Marx}},\ and\ \bibinfo {author} {\bibfnamefont {M.}~\bibnamefont {Parrinello}},\ }\bibfield  {title} {\bibinfo {title} {Tunnelling and zero-point motion in high-pressure ice},\ }\href@noop {} {\bibfield  {journal} {\bibinfo  {journal} {Nature}\ }\textbf {\bibinfo {volume} {392}},\ \bibinfo {pages} {258} (\bibinfo {year} {1998})}\BibitemShut {NoStop}%
\bibitem [{\citenamefont {Kang}\ \emph {et~al.}(2013)\citenamefont {Kang}, \citenamefont {Dai}, \citenamefont {Sun}, \citenamefont {Hou},\ and\ \citenamefont {Yuan}}]{kang2013quantum}%
  \BibitemOpen
  \bibfield  {author} {\bibinfo {author} {\bibfnamefont {D.}~\bibnamefont {Kang}}, \bibinfo {author} {\bibfnamefont {J.}~\bibnamefont {Dai}}, \bibinfo {author} {\bibfnamefont {H.}~\bibnamefont {Sun}}, \bibinfo {author} {\bibfnamefont {Y.}~\bibnamefont {Hou}},\ and\ \bibinfo {author} {\bibfnamefont {J.}~\bibnamefont {Yuan}},\ }\bibfield  {title} {\bibinfo {title} {Quantum simulation of thermally-driven phase transition and oxygen k-edge x-ray absorption of high-pressure ice},\ }\href@noop {} {\bibfield  {journal} {\bibinfo  {journal} {Scientific reports}\ }\textbf {\bibinfo {volume} {3}},\ \bibinfo {pages} {3272} (\bibinfo {year} {2013})}\BibitemShut {NoStop}%
\bibitem [{\citenamefont {Cherubini}\ \emph {et~al.}(2024)\citenamefont {Cherubini}, \citenamefont {Monacelli}, \citenamefont {Yang}, \citenamefont {Car}, \citenamefont {Casula},\ and\ \citenamefont {Mauri}}]{cherubini2024quantum}%
  \BibitemOpen
  \bibfield  {author} {\bibinfo {author} {\bibfnamefont {M.}~\bibnamefont {Cherubini}}, \bibinfo {author} {\bibfnamefont {L.}~\bibnamefont {Monacelli}}, \bibinfo {author} {\bibfnamefont {B.}~\bibnamefont {Yang}}, \bibinfo {author} {\bibfnamefont {R.}~\bibnamefont {Car}}, \bibinfo {author} {\bibfnamefont {M.}~\bibnamefont {Casula}},\ and\ \bibinfo {author} {\bibfnamefont {F.}~\bibnamefont {Mauri}},\ }\bibfield  {title} {\bibinfo {title} {Quantum effects in h-bond symmetrization and in thermodynamic properties of high pressure ice},\ }\href@noop {} {\bibfield  {journal} {\bibinfo  {journal} {Physical Review B}\ }\textbf {\bibinfo {volume} {110}},\ \bibinfo {pages} {014112} (\bibinfo {year} {2024})}\BibitemShut {NoStop}%
\bibitem [{\citenamefont {Benoit}\ \emph {et~al.}(1996)\citenamefont {Benoit}, \citenamefont {Bernasconi}, \citenamefont {Focher},\ and\ \citenamefont {Parrinello}}]{benoit1996new}%
  \BibitemOpen
  \bibfield  {author} {\bibinfo {author} {\bibfnamefont {M.}~\bibnamefont {Benoit}}, \bibinfo {author} {\bibfnamefont {M.}~\bibnamefont {Bernasconi}}, \bibinfo {author} {\bibfnamefont {P.}~\bibnamefont {Focher}},\ and\ \bibinfo {author} {\bibfnamefont {M.}~\bibnamefont {Parrinello}},\ }\bibfield  {title} {\bibinfo {title} {New high-pressure phase of ice},\ }\href@noop {} {\bibfield  {journal} {\bibinfo  {journal} {Physical review letters}\ }\textbf {\bibinfo {volume} {76}},\ \bibinfo {pages} {2934} (\bibinfo {year} {1996})}\BibitemShut {NoStop}%
\bibitem [{\citenamefont {Pinsook}\ and\ \citenamefont {Ackland}(1998)}]{pinsook1998simulation}%
  \BibitemOpen
  \bibfield  {author} {\bibinfo {author} {\bibfnamefont {U.}~\bibnamefont {Pinsook}}\ and\ \bibinfo {author} {\bibfnamefont {G.~J.}\ \bibnamefont {Ackland}},\ }\bibfield  {title} {\bibinfo {title} {Simulation of martensitic microstructural evolution in zirconium},\ }\href@noop {} {\bibfield  {journal} {\bibinfo  {journal} {Physical Review B}\ }\textbf {\bibinfo {volume} {58}},\ \bibinfo {pages} {11252} (\bibinfo {year} {1998})}\BibitemShut {NoStop}%
\bibitem [{\citenamefont {Pinsook}\ and\ \citenamefont {Ackland}(2000)}]{pinsook2000atomistic}%
  \BibitemOpen
  \bibfield  {author} {\bibinfo {author} {\bibfnamefont {U.}~\bibnamefont {Pinsook}}\ and\ \bibinfo {author} {\bibfnamefont {G.~J.}\ \bibnamefont {Ackland}},\ }\bibfield  {title} {\bibinfo {title} {Atomistic simulation of shear in a martensitic twinned microstructure},\ }\href@noop {} {\bibfield  {journal} {\bibinfo  {journal} {Physical Review B}\ }\textbf {\bibinfo {volume} {62}},\ \bibinfo {pages} {5427} (\bibinfo {year} {2000})}\BibitemShut {NoStop}%
\bibitem [{\citenamefont {Pinsook}\ and\ \citenamefont {Ackland}(1999)}]{pinsook1999calculation}%
  \BibitemOpen
  \bibfield  {author} {\bibinfo {author} {\bibfnamefont {U.}~\bibnamefont {Pinsook}}\ and\ \bibinfo {author} {\bibfnamefont {G.~J.}\ \bibnamefont {Ackland}},\ }\bibfield  {title} {\bibinfo {title} {Calculation of anomalous phonons and the hcp-bcc phase transition in zirconium},\ }\href@noop {} {\bibfield  {journal} {\bibinfo  {journal} {Physical Review B}\ }\textbf {\bibinfo {volume} {59}},\ \bibinfo {pages} {13642} (\bibinfo {year} {1999})}\BibitemShut {NoStop}%
\bibitem [{\citenamefont {Weck}\ \emph {et~al.}(2022)\citenamefont {Weck}, \citenamefont {Queyroux}, \citenamefont {Ninet}, \citenamefont {Datchi}, \citenamefont {Mezouar},\ and\ \citenamefont {Loubeyre}}]{weck2022evidence}%
  \BibitemOpen
  \bibfield  {author} {\bibinfo {author} {\bibfnamefont {G.}~\bibnamefont {Weck}}, \bibinfo {author} {\bibfnamefont {J.-A.}\ \bibnamefont {Queyroux}}, \bibinfo {author} {\bibfnamefont {S.}~\bibnamefont {Ninet}}, \bibinfo {author} {\bibfnamefont {F.}~\bibnamefont {Datchi}}, \bibinfo {author} {\bibfnamefont {M.}~\bibnamefont {Mezouar}},\ and\ \bibinfo {author} {\bibfnamefont {P.}~\bibnamefont {Loubeyre}},\ }\bibfield  {title} {\bibinfo {title} {Evidence and stability field of fcc superionic water ice using static compression},\ }\href@noop {} {\bibfield  {journal} {\bibinfo  {journal} {Physical Review Letters}\ }\textbf {\bibinfo {volume} {128}},\ \bibinfo {pages} {165701} (\bibinfo {year} {2022})}\BibitemShut {NoStop}%
\bibitem [{\citenamefont {Millot}\ \emph {et~al.}(2019)\citenamefont {Millot}, \citenamefont {Coppari}, \citenamefont {Rygg}, \citenamefont {Correa~Barrios}, \citenamefont {Hamel}, \citenamefont {Swift},\ and\ \citenamefont {Eggert}}]{millot2019nanosecond}%
  \BibitemOpen
  \bibfield  {author} {\bibinfo {author} {\bibfnamefont {M.}~\bibnamefont {Millot}}, \bibinfo {author} {\bibfnamefont {F.}~\bibnamefont {Coppari}}, \bibinfo {author} {\bibfnamefont {J.~R.}\ \bibnamefont {Rygg}}, \bibinfo {author} {\bibfnamefont {A.}~\bibnamefont {Correa~Barrios}}, \bibinfo {author} {\bibfnamefont {S.}~\bibnamefont {Hamel}}, \bibinfo {author} {\bibfnamefont {D.~C.}\ \bibnamefont {Swift}},\ and\ \bibinfo {author} {\bibfnamefont {J.~H.}\ \bibnamefont {Eggert}},\ }\bibfield  {title} {\bibinfo {title} {Nanosecond x-ray diffraction of shock-compressed superionic water ice},\ }\href@noop {} {\bibfield  {journal} {\bibinfo  {journal} {Nature}\ }\textbf {\bibinfo {volume} {569}},\ \bibinfo {pages} {251} (\bibinfo {year} {2019})}\BibitemShut {NoStop}%
\bibitem [{\citenamefont {Cavazzoni}\ \emph {et~al.}(1999)\citenamefont {Cavazzoni}, \citenamefont {Chiarotti}, \citenamefont {Scandolo}, \citenamefont {Tosatti}, \citenamefont {Bernasconi},\ and\ \citenamefont {Parrinello}}]{cavazzoni1999superionic}%
  \BibitemOpen
  \bibfield  {author} {\bibinfo {author} {\bibfnamefont {C.}~\bibnamefont {Cavazzoni}}, \bibinfo {author} {\bibfnamefont {G.}~\bibnamefont {Chiarotti}}, \bibinfo {author} {\bibfnamefont {S.}~\bibnamefont {Scandolo}}, \bibinfo {author} {\bibfnamefont {E.}~\bibnamefont {Tosatti}}, \bibinfo {author} {\bibfnamefont {M.}~\bibnamefont {Bernasconi}},\ and\ \bibinfo {author} {\bibfnamefont {M.}~\bibnamefont {Parrinello}},\ }\bibfield  {title} {\bibinfo {title} {Superionic and metallic states of water and ammonia at giant planet conditions},\ }\href@noop {} {\bibfield  {journal} {\bibinfo  {journal} {Science}\ }\textbf {\bibinfo {volume} {283}},\ \bibinfo {pages} {44} (\bibinfo {year} {1999})}\BibitemShut {NoStop}%
\bibitem [{\citenamefont {Prakapenka}\ \emph {et~al.}(2021)\citenamefont {Prakapenka}, \citenamefont {Holtgrewe}, \citenamefont {Lobanov},\ and\ \citenamefont {Goncharov}}]{prakapenka2021structure}%
  \BibitemOpen
  \bibfield  {author} {\bibinfo {author} {\bibfnamefont {V.~B.}\ \bibnamefont {Prakapenka}}, \bibinfo {author} {\bibfnamefont {N.}~\bibnamefont {Holtgrewe}}, \bibinfo {author} {\bibfnamefont {S.~S.}\ \bibnamefont {Lobanov}},\ and\ \bibinfo {author} {\bibfnamefont {A.~F.}\ \bibnamefont {Goncharov}},\ }\bibfield  {title} {\bibinfo {title} {Structure and properties of two superionic ice phases},\ }\href@noop {} {\bibfield  {journal} {\bibinfo  {journal} {Nature Physics}\ }\textbf {\bibinfo {volume} {17}},\ \bibinfo {pages} {1233} (\bibinfo {year} {2021})}\BibitemShut {NoStop}%
\bibitem [{\citenamefont {Rescigno}\ \emph {et~al.}(2025)\citenamefont {Rescigno}, \citenamefont {Toffano}, \citenamefont {Ranieri}, \citenamefont {Andriambariarijaona}, \citenamefont {Gaal}, \citenamefont {Klotz}, \citenamefont {Koza}, \citenamefont {Ollivier}, \citenamefont {Martelli}, \citenamefont {Russo} \emph {et~al.}}]{rescigno2025observation}%
  \BibitemOpen
  \bibfield  {author} {\bibinfo {author} {\bibfnamefont {M.}~\bibnamefont {Rescigno}}, \bibinfo {author} {\bibfnamefont {A.}~\bibnamefont {Toffano}}, \bibinfo {author} {\bibfnamefont {U.}~\bibnamefont {Ranieri}}, \bibinfo {author} {\bibfnamefont {L.}~\bibnamefont {Andriambariarijaona}}, \bibinfo {author} {\bibfnamefont {R.}~\bibnamefont {Gaal}}, \bibinfo {author} {\bibfnamefont {S.}~\bibnamefont {Klotz}}, \bibinfo {author} {\bibfnamefont {M.~M.}\ \bibnamefont {Koza}}, \bibinfo {author} {\bibfnamefont {J.}~\bibnamefont {Ollivier}}, \bibinfo {author} {\bibfnamefont {F.}~\bibnamefont {Martelli}}, \bibinfo {author} {\bibfnamefont {J.}~\bibnamefont {Russo}}, \emph {et~al.},\ }\bibfield  {title} {\bibinfo {title} {Observation of plastic ice vii by quasi-elastic neutron scattering},\ }\href@noop {} {\bibfield  {journal} {\bibinfo  {journal} {Nature}\ ,\ \bibinfo {pages} {1}} (\bibinfo {year} {2025})}\BibitemShut {NoStop}%
\bibitem [{\citenamefont {Minceva-Sukarova}\ \emph {et~al.}(1984)\citenamefont {Minceva-Sukarova}, \citenamefont {Sherman},\ and\ \citenamefont {Wilkinson}}]{minceva1984}%
  \BibitemOpen
  \bibfield  {author} {\bibinfo {author} {\bibfnamefont {B.}~\bibnamefont {Minceva-Sukarova}}, \bibinfo {author} {\bibfnamefont {W.~F.}\ \bibnamefont {Sherman}},\ and\ \bibinfo {author} {\bibfnamefont {G.~R.}\ \bibnamefont {Wilkinson}},\ }\bibfield  {title} {\bibinfo {title} {The {Raman} spectra of ice ({Ih}, {II}, {III}, {V}, {VI} and {IX}) as functions of pressure and temperature},\ }\href {https://doi.org/10.1088/0022-3719/17/32/017} {\bibfield  {journal} {\bibinfo  {journal} {Journal of Physics C: Solid State Physics}\ }\textbf {\bibinfo {volume} {17}},\ \bibinfo {pages} {5833} (\bibinfo {year} {1984})}\BibitemShut {NoStop}%
\bibitem [{\citenamefont {Pruzan}\ \emph {et~al.}(1990)\citenamefont {Pruzan}, \citenamefont {Chervin},\ and\ \citenamefont {Gauthier}}]{pruzan1990raman}%
  \BibitemOpen
  \bibfield  {author} {\bibinfo {author} {\bibfnamefont {P.}~\bibnamefont {Pruzan}}, \bibinfo {author} {\bibfnamefont {J.}~\bibnamefont {Chervin}},\ and\ \bibinfo {author} {\bibfnamefont {M.}~\bibnamefont {Gauthier}},\ }\bibfield  {title} {\bibinfo {title} {Raman spectroscopy investigation of ice vii and deuterated ice vii to 40 gpa. disorder in ice vii},\ }\href@noop {} {\bibfield  {journal} {\bibinfo  {journal} {Europhysics Letters}\ }\textbf {\bibinfo {volume} {13}},\ \bibinfo {pages} {81} (\bibinfo {year} {1990})}\BibitemShut {NoStop}%
\bibitem [{\citenamefont {Li}\ \emph {et~al.}(2024)\citenamefont {Li}, \citenamefont {Gao}, \citenamefont {Zhang}, \citenamefont {Liu}, \citenamefont {Lv},\ and\ \citenamefont {Wang}}]{li2024phase}%
  \BibitemOpen
  \bibfield  {author} {\bibinfo {author} {\bibfnamefont {L.}~\bibnamefont {Li}}, \bibinfo {author} {\bibfnamefont {P.}~\bibnamefont {Gao}}, \bibinfo {author} {\bibfnamefont {W.}~\bibnamefont {Zhang}}, \bibinfo {author} {\bibfnamefont {G.}~\bibnamefont {Liu}}, \bibinfo {author} {\bibfnamefont {J.}~\bibnamefont {Lv}},\ and\ \bibinfo {author} {\bibfnamefont {Y.}~\bibnamefont {Wang}},\ }\bibfield  {title} {\bibinfo {title} {Phase transitions of ice viii-vii-x: A potential energy landscape perspective},\ }\href@noop {} {\bibfield  {journal} {\bibinfo  {journal} {Physical Review Research}\ }\textbf {\bibinfo {volume} {6}},\ \bibinfo {pages} {043085} (\bibinfo {year} {2024})}\BibitemShut {NoStop}%
\bibitem [{\citenamefont {Umemoto}\ \emph {et~al.}(2010)\citenamefont {Umemoto}, \citenamefont {Wentzcovitch}, \citenamefont {De~Gironcoli},\ and\ \citenamefont {Baroni}}]{umemoto2010order}%
  \BibitemOpen
  \bibfield  {author} {\bibinfo {author} {\bibfnamefont {K.}~\bibnamefont {Umemoto}}, \bibinfo {author} {\bibfnamefont {R.~M.}\ \bibnamefont {Wentzcovitch}}, \bibinfo {author} {\bibfnamefont {S.}~\bibnamefont {De~Gironcoli}},\ and\ \bibinfo {author} {\bibfnamefont {S.}~\bibnamefont {Baroni}},\ }\bibfield  {title} {\bibinfo {title} {Order--disorder phase boundary between ice vii and viii obtained by first principles},\ }\href@noop {} {\bibfield  {journal} {\bibinfo  {journal} {Chemical Physics Letters}\ }\textbf {\bibinfo {volume} {499}},\ \bibinfo {pages} {236} (\bibinfo {year} {2010})}\BibitemShut {NoStop}%
\bibitem [{\citenamefont {Bjerrum}(1952)}]{bjerrum1952structure}%
  \BibitemOpen
  \bibfield  {author} {\bibinfo {author} {\bibfnamefont {N.}~\bibnamefont {Bjerrum}},\ }\bibfield  {title} {\bibinfo {title} {Structure and properties of ice},\ }\href@noop {} {\bibfield  {journal} {\bibinfo  {journal} {Science}\ }\textbf {\bibinfo {volume} {115}},\ \bibinfo {pages} {385} (\bibinfo {year} {1952})}\BibitemShut {NoStop}%
\bibitem [{Note3()}]{Note3}%
  \BibitemOpen
  \bibinfo {note} {Notice that the physics here is similar to forming a chemical bond, except that here it is the proton rather than the electron shared between two atoms}\BibitemShut {NoStop}%
\bibitem [{\citenamefont {Falk}\ and\ \citenamefont {Gigu{\`e}re}(1957)}]{falk1957infrared}%
  \BibitemOpen
  \bibfield  {author} {\bibinfo {author} {\bibfnamefont {M.}~\bibnamefont {Falk}}\ and\ \bibinfo {author} {\bibfnamefont {P.~A.}\ \bibnamefont {Gigu{\`e}re}},\ }\bibfield  {title} {\bibinfo {title} {Infrared spectrum of the h3o+ ion in aqueous solutions},\ }\href@noop {} {\bibfield  {journal} {\bibinfo  {journal} {Canadian Journal of Chemistry}\ }\textbf {\bibinfo {volume} {35}},\ \bibinfo {pages} {1195} (\bibinfo {year} {1957})}\BibitemShut {NoStop}%
\bibitem [{\citenamefont {Cooke}\ \emph {et~al.}(2020)\citenamefont {Cooke}, \citenamefont {Magd{\u{a}}u}, \citenamefont {Pena-Alvarez}, \citenamefont {Afonina}, \citenamefont {Dalladay-Simpson}, \citenamefont {Liu}, \citenamefont {Howie}, \citenamefont {Gregoryanz},\ and\ \citenamefont {Ackland}}]{cooke2020raman}%
  \BibitemOpen
  \bibfield  {author} {\bibinfo {author} {\bibfnamefont {P.~I.}\ \bibnamefont {Cooke}}, \bibinfo {author} {\bibfnamefont {I.~B.}\ \bibnamefont {Magd{\u{a}}u}}, \bibinfo {author} {\bibfnamefont {M.}~\bibnamefont {Pena-Alvarez}}, \bibinfo {author} {\bibfnamefont {V.}~\bibnamefont {Afonina}}, \bibinfo {author} {\bibfnamefont {P.}~\bibnamefont {Dalladay-Simpson}}, \bibinfo {author} {\bibfnamefont {X.-D.}\ \bibnamefont {Liu}}, \bibinfo {author} {\bibfnamefont {R.~T.}\ \bibnamefont {Howie}}, \bibinfo {author} {\bibfnamefont {E.}~\bibnamefont {Gregoryanz}},\ and\ \bibinfo {author} {\bibfnamefont {G.~J.}\ \bibnamefont {Ackland}},\ }\bibfield  {title} {\bibinfo {title} {Raman signal from a hindered hydrogen rotor},\ }\href@noop {} {\bibfield  {journal} {\bibinfo  {journal} {Physical Review B}\ }\textbf {\bibinfo {volume} {102}},\ \bibinfo {pages} {064102} (\bibinfo {year} {2020})}\BibitemShut {NoStop}%
\bibitem [{\citenamefont {Cooke}\ \emph {et~al.}(2022)\citenamefont {Cooke}, \citenamefont {Magd{\u{a}}u},\ and\ \citenamefont {Ackland}}]{cooke2022calculating}%
  \BibitemOpen
  \bibfield  {author} {\bibinfo {author} {\bibfnamefont {P.~I.}\ \bibnamefont {Cooke}}, \bibinfo {author} {\bibfnamefont {I.~B.}\ \bibnamefont {Magd{\u{a}}u}},\ and\ \bibinfo {author} {\bibfnamefont {G.~J.}\ \bibnamefont {Ackland}},\ }\bibfield  {title} {\bibinfo {title} {Calculating the raman signal beyond perturbation theory for a diatomic molecular crystal},\ }\href@noop {} {\bibfield  {journal} {\bibinfo  {journal} {Computational Materials Science}\ }\textbf {\bibinfo {volume} {210}},\ \bibinfo {pages} {111400} (\bibinfo {year} {2022})}\BibitemShut {NoStop}%
\bibitem [{\citenamefont {Ramsey}\ \emph {et~al.}(2020)\citenamefont {Ramsey}, \citenamefont {Pena-Alvarez},\ and\ \citenamefont {Ackland}}]{ramsey2020localization}%
  \BibitemOpen
  \bibfield  {author} {\bibinfo {author} {\bibfnamefont {S.~B.}\ \bibnamefont {Ramsey}}, \bibinfo {author} {\bibfnamefont {M.}~\bibnamefont {Pena-Alvarez}},\ and\ \bibinfo {author} {\bibfnamefont {G.~J.}\ \bibnamefont {Ackland}},\ }\bibfield  {title} {\bibinfo {title} {Localization effects on the vibron shifts in helium-hydrogen mixtures},\ }\href@noop {} {\bibfield  {journal} {\bibinfo  {journal} {Physical Review B}\ }\textbf {\bibinfo {volume} {101}},\ \bibinfo {pages} {214306} (\bibinfo {year} {2020})}\BibitemShut {NoStop}%
\bibitem [{\citenamefont {Magd{\u{a}}u}\ and\ \citenamefont {Ackland}(2013)}]{magduau2013identification}%
  \BibitemOpen
  \bibfield  {author} {\bibinfo {author} {\bibfnamefont {I.~B.}\ \bibnamefont {Magd{\u{a}}u}}\ and\ \bibinfo {author} {\bibfnamefont {G.~J.}\ \bibnamefont {Ackland}},\ }\bibfield  {title} {\bibinfo {title} {Identification of high-pressure phases iii and iv in hydrogen: Simulating raman spectra using molecular dynamics},\ }\href@noop {} {\bibfield  {journal} {\bibinfo  {journal} {Physical Review B—Condensed Matter and Materials Physics}\ }\textbf {\bibinfo {volume} {87}},\ \bibinfo {pages} {174110} (\bibinfo {year} {2013})}\BibitemShut {NoStop}%
\bibitem [{\citenamefont {Pe{\~n}a-Alvarez}\ \emph {et~al.}(2020)\citenamefont {Pe{\~n}a-Alvarez}, \citenamefont {Afonina}, \citenamefont {Dalladay-Simpson}, \citenamefont {Liu}, \citenamefont {Howie}, \citenamefont {Cooke}, \citenamefont {Magdau}, \citenamefont {Ackland},\ and\ \citenamefont {Gregoryanz}}]{pena2020quantitative}%
  \BibitemOpen
  \bibfield  {author} {\bibinfo {author} {\bibfnamefont {M.}~\bibnamefont {Pe{\~n}a-Alvarez}}, \bibinfo {author} {\bibfnamefont {V.}~\bibnamefont {Afonina}}, \bibinfo {author} {\bibfnamefont {P.}~\bibnamefont {Dalladay-Simpson}}, \bibinfo {author} {\bibfnamefont {X.-D.}\ \bibnamefont {Liu}}, \bibinfo {author} {\bibfnamefont {R.~T.}\ \bibnamefont {Howie}}, \bibinfo {author} {\bibfnamefont {P.~I.}\ \bibnamefont {Cooke}}, \bibinfo {author} {\bibfnamefont {I.~B.}\ \bibnamefont {Magdau}}, \bibinfo {author} {\bibfnamefont {G.~J.}\ \bibnamefont {Ackland}},\ and\ \bibinfo {author} {\bibfnamefont {E.}~\bibnamefont {Gregoryanz}},\ }\bibfield  {title} {\bibinfo {title} {Quantitative rotational to librational transition in dense h2 and d2},\ }\href@noop {} {\bibfield  {journal} {\bibinfo  {journal} {The Journal of Physical Chemistry Letters}\ }\textbf {\bibinfo {volume} {11}},\ \bibinfo {pages} {6626} (\bibinfo {year} {2020})}\BibitemShut {NoStop}%
\bibitem [{\citenamefont {Howie}\ \emph {et~al.}(2014)\citenamefont {Howie}, \citenamefont {Magd{\u{a}}u}, \citenamefont {Goncharov}, \citenamefont {Ackland},\ and\ \citenamefont {Gregoryanz}}]{howie2014phonon}%
  \BibitemOpen
  \bibfield  {author} {\bibinfo {author} {\bibfnamefont {R.~T.}\ \bibnamefont {Howie}}, \bibinfo {author} {\bibfnamefont {I.~B.}\ \bibnamefont {Magd{\u{a}}u}}, \bibinfo {author} {\bibfnamefont {A.~F.}\ \bibnamefont {Goncharov}}, \bibinfo {author} {\bibfnamefont {G.~J.}\ \bibnamefont {Ackland}},\ and\ \bibinfo {author} {\bibfnamefont {E.}~\bibnamefont {Gregoryanz}},\ }\bibfield  {title} {\bibinfo {title} {Phonon localization by mass disorder in dense hydrogen-deuterium binary alloy},\ }\href@noop {} {\bibfield  {journal} {\bibinfo  {journal} {Physical review letters}\ }\textbf {\bibinfo {volume} {113}},\ \bibinfo {pages} {175501} (\bibinfo {year} {2014})}\BibitemShut {NoStop}%
\bibitem [{\citenamefont {Frost}\ \emph {et~al.}(2023)\citenamefont {Frost}, \citenamefont {Hermann}, \citenamefont {Glenzer},\ and\ \citenamefont {Ackland}}]{frost2023isotopic}%
  \BibitemOpen
  \bibfield  {author} {\bibinfo {author} {\bibfnamefont {M.}~\bibnamefont {Frost}}, \bibinfo {author} {\bibfnamefont {A.}~\bibnamefont {Hermann}}, \bibinfo {author} {\bibfnamefont {S.~H.}\ \bibnamefont {Glenzer}},\ and\ \bibinfo {author} {\bibfnamefont {G.~J.}\ \bibnamefont {Ackland}},\ }\bibfield  {title} {\bibinfo {title} {Isotopic raman broadening due to anderson localization of harmonic phonons in partially deuterated ice vii and viii},\ }\href@noop {} {\bibfield  {journal} {\bibinfo  {journal} {Physical Review Research}\ }\textbf {\bibinfo {volume} {5}},\ \bibinfo {pages} {043166} (\bibinfo {year} {2023})}\BibitemShut {NoStop}%
\bibitem [{\citenamefont {Ackland}\ and\ \citenamefont {Loveday}(2020)}]{ackland2020structures}%
  \BibitemOpen
  \bibfield  {author} {\bibinfo {author} {\bibfnamefont {G.~J.}\ \bibnamefont {Ackland}}\ and\ \bibinfo {author} {\bibfnamefont {J.~S.}\ \bibnamefont {Loveday}},\ }\bibfield  {title} {\bibinfo {title} {Structures of solid hydrogen at 300 k},\ }\href@noop {} {\bibfield  {journal} {\bibinfo  {journal} {Physical Review B}\ }\textbf {\bibinfo {volume} {101}},\ \bibinfo {pages} {094104} (\bibinfo {year} {2020})}\BibitemShut {NoStop}%
\bibitem [{\citenamefont {Magd{\u{a}}u}\ and\ \citenamefont {Ackland}(2017)}]{magduau2017infrared}%
  \BibitemOpen
  \bibfield  {author} {\bibinfo {author} {\bibfnamefont {I.~B.}\ \bibnamefont {Magd{\u{a}}u}}\ and\ \bibinfo {author} {\bibfnamefont {G.~J.}\ \bibnamefont {Ackland}},\ }\bibfield  {title} {\bibinfo {title} {Infrared peak splitting from phonon localization in solid hydrogen},\ }\href@noop {} {\bibfield  {journal} {\bibinfo  {journal} {Physical review letters}\ }\textbf {\bibinfo {volume} {118}},\ \bibinfo {pages} {145701} (\bibinfo {year} {2017})}\BibitemShut {NoStop}%
\bibitem [{\citenamefont {Komatsu}\ \emph {et~al.}(2020)\citenamefont {Komatsu}, \citenamefont {Klotz}, \citenamefont {Machida}, \citenamefont {Sano-Furukawa}, \citenamefont {Hattori},\ and\ \citenamefont {Kagi}}]{komatsu2020anomalous}%
  \BibitemOpen
  \bibfield  {author} {\bibinfo {author} {\bibfnamefont {K.}~\bibnamefont {Komatsu}}, \bibinfo {author} {\bibfnamefont {S.}~\bibnamefont {Klotz}}, \bibinfo {author} {\bibfnamefont {S.}~\bibnamefont {Machida}}, \bibinfo {author} {\bibfnamefont {A.}~\bibnamefont {Sano-Furukawa}}, \bibinfo {author} {\bibfnamefont {T.}~\bibnamefont {Hattori}},\ and\ \bibinfo {author} {\bibfnamefont {H.}~\bibnamefont {Kagi}},\ }\bibfield  {title} {\bibinfo {title} {Anomalous hydrogen dynamics of the ice vii--viii transition revealed by high-pressure neutron diffraction},\ }\href@noop {} {\bibfield  {journal} {\bibinfo  {journal} {Proceedings of the National Academy of Sciences}\ }\textbf {\bibinfo {volume} {117}},\ \bibinfo {pages} {6356} (\bibinfo {year} {2020})}\BibitemShut {NoStop}%
\bibitem [{\citenamefont {Jorgensen}\ and\ \citenamefont {Worlton}(1985)}]{jorgensen1985}%
  \BibitemOpen
  \bibfield  {author} {\bibinfo {author} {\bibfnamefont {J.~D.}\ \bibnamefont {Jorgensen}}\ and\ \bibinfo {author} {\bibfnamefont {T.~G.}\ \bibnamefont {Worlton}},\ }\bibfield  {title} {\bibinfo {title} {Disordered structure of {D2O} ice {VII} from in situ neutron powder diffraction},\ }\href {https://doi.org/10.1063/1.449867} {\bibfield  {journal} {\bibinfo  {journal} {The Journal of Chemical Physics}\ }\textbf {\bibinfo {volume} {83}},\ \bibinfo {pages} {329} (\bibinfo {year} {1985})}\BibitemShut {NoStop}%
\bibitem [{\citenamefont {Nelmes}\ \emph {et~al.}(1998)\citenamefont {Nelmes}, \citenamefont {Loveday}, \citenamefont {Marshall}, \citenamefont {Hamel},\ and\ \citenamefont {Besson}}]{nelmes1998}%
  \BibitemOpen
  \bibfield  {author} {\bibinfo {author} {\bibfnamefont {R.}~\bibnamefont {Nelmes}}, \bibinfo {author} {\bibfnamefont {J.}~\bibnamefont {Loveday}}, \bibinfo {author} {\bibfnamefont {W.}~\bibnamefont {Marshall}}, \bibinfo {author} {\bibfnamefont {G.}~\bibnamefont {Hamel}},\ and\ \bibinfo {author} {\bibfnamefont {J.}~\bibnamefont {Besson}},\ }\bibfield  {title} {\bibinfo {title} {Multisite {Disordered} {Structure} of {Ice} {VII} to 20 {GPa}},\ }\bibfield  {journal} {\bibinfo  {journal} {Physical Review Letters}\ }\textbf {\bibinfo {volume} {81}},\ \href {https://doi.org/10.1103/PhysRevLett.81.2719} {10.1103/PhysRevLett.81.2719} (\bibinfo {year} {1998})\BibitemShut {NoStop}%
\bibitem [{\citenamefont {Yamashita}\ \emph {et~al.}(2022)\citenamefont {Yamashita}, \citenamefont {Komatsu}, \citenamefont {Klotz}, \citenamefont {Fabelo}, \citenamefont {Fernández-Díaz}, \citenamefont {Abe}, \citenamefont {Machida}, \citenamefont {Hattori}, \citenamefont {Irifune}, \citenamefont {Shinmei}, \citenamefont {Sugiyama}, \citenamefont {Kawamata},\ and\ \citenamefont {Kagi}}]{yamashita2022}%
  \BibitemOpen
  \bibfield  {author} {\bibinfo {author} {\bibfnamefont {K.}~\bibnamefont {Yamashita}}, \bibinfo {author} {\bibfnamefont {K.}~\bibnamefont {Komatsu}}, \bibinfo {author} {\bibfnamefont {S.}~\bibnamefont {Klotz}}, \bibinfo {author} {\bibfnamefont {O.}~\bibnamefont {Fabelo}}, \bibinfo {author} {\bibfnamefont {M.~T.}\ \bibnamefont {Fernández-Díaz}}, \bibinfo {author} {\bibfnamefont {J.}~\bibnamefont {Abe}}, \bibinfo {author} {\bibfnamefont {S.}~\bibnamefont {Machida}}, \bibinfo {author} {\bibfnamefont {T.}~\bibnamefont {Hattori}}, \bibinfo {author} {\bibfnamefont {T.}~\bibnamefont {Irifune}}, \bibinfo {author} {\bibfnamefont {T.}~\bibnamefont {Shinmei}}, \bibinfo {author} {\bibfnamefont {K.}~\bibnamefont {Sugiyama}}, \bibinfo {author} {\bibfnamefont {T.}~\bibnamefont {Kawamata}},\ and\ \bibinfo {author} {\bibfnamefont {H.}~\bibnamefont {Kagi}},\ }\bibfield  {title} {\bibinfo {title} {Atomic distribution and local structure in ice {VII} from in situ neutron diffraction},\ }\href
  {https://doi.org/10.1073/pnas.2208717119} {\bibfield  {journal} {\bibinfo  {journal} {Proceedings of the National Academy of Sciences}\ }\textbf {\bibinfo {volume} {119}},\ \bibinfo {pages} {e2208717119} (\bibinfo {year} {2022})},\ \bibinfo {note} {publisher: Proceedings of the National Academy of Sciences}\BibitemShut {NoStop}%
\bibitem [{\citenamefont {Singer}\ \emph {et~al.}(2005)\citenamefont {Singer}, \citenamefont {Kuo}, \citenamefont {Hirsch}, \citenamefont {Knight}, \citenamefont {Ojam{\"a}e},\ and\ \citenamefont {Klein}}]{singer2005hydrogen}%
  \BibitemOpen
  \bibfield  {author} {\bibinfo {author} {\bibfnamefont {S.~J.}\ \bibnamefont {Singer}}, \bibinfo {author} {\bibfnamefont {J.-L.}\ \bibnamefont {Kuo}}, \bibinfo {author} {\bibfnamefont {T.~K.}\ \bibnamefont {Hirsch}}, \bibinfo {author} {\bibfnamefont {C.}~\bibnamefont {Knight}}, \bibinfo {author} {\bibfnamefont {L.}~\bibnamefont {Ojam{\"a}e}},\ and\ \bibinfo {author} {\bibfnamefont {M.~L.}\ \bibnamefont {Klein}},\ }\bibfield  {title} {\bibinfo {title} {Hydrogen-bond topology and the ice vii/viii and ice i h/xi<? format?> proton-ordering phase transitions},\ }\href@noop {} {\bibfield  {journal} {\bibinfo  {journal} {Physical review letters}\ }\textbf {\bibinfo {volume} {94}},\ \bibinfo {pages} {135701} (\bibinfo {year} {2005})}\BibitemShut {NoStop}%
\end{thebibliography}%

\end{document}